\documentclass[journal]{IEEEtran}
\usepackage{amsmath,amssymb}
\usepackage{amsthm}
\usepackage{array}
\usepackage[caption=false,font=normalsize,labelfont=sf,textfont=sf]{subfig}
\usepackage{textcomp}
\usepackage{stfloats}
\usepackage{url}
\usepackage{verbatim}
\usepackage{graphicx}
\usepackage{cite}
\usepackage{bm}
\usepackage{enumitem} 
\usepackage{multirow}
\usepackage[linesnumbered,ruled,vlined]{algorithm2e}
\usepackage{booktabs}
\usepackage{xcolor}

\allowdisplaybreaks

\title{Adaptive Privacy of Sequential Data Releases Under Collusion}
\author{Sophie~Taylor, Praneeth~Kumar~Vippathalla, and Justin~P.~Coon 
\thanks{The authors are with the Department of Engineering Science, University of Oxford, Parks Road, Oxford, OX1 3PJ, UK,
(email: \protect\url{sophie.taylor2@balliol.ox.ac.uk}; \protect\url{praneeth.vippathalla@eng.ox.ac.uk}; \protect\url{justin.coon@eng.ox.ac.uk)}.}
\thanks{This research was funded in part by the Engineering and Physical Sciences Research Council under grant number EP/W524311/1, and the U. S. Army Research Laboratory and the U. S. Army Research Office under grant number W911NF-22-1-0070. For the purpose of Open Access, the authors have applied a CC BY public copyright license to any Author Accepted Manuscript (AAM) version arising from this submission.}}
\date{}

\begin{document}
\maketitle

\begin{abstract}
    The fundamental trade-off between privacy and utility remains an active area of research.
    Our contribution is motivated by two observations.
    First, privacy mechanisms developed for one-time data release cannot straightforwardly be extended to sequential releases.
    Second, practical databases are likely to be useful to multiple distinct parties.
    Furthermore, we can not rule out the possibility of data sharing between parties.
    With utility in mind, we formulate a new privacy-utility trade-off problem to adaptively tackle sequential data requests made by different, potentially colluding entities.
    We consider both expected distortion and mutual information as measures to quantify utility,
    and use mutual information to measure privacy.
    We
    assume an attack model whereby illicit data sharing, which we call collusion, can occur between data receivers.
    We develop an adaptive algorithm for data releases that makes use of a Blahut--Arimoto-style algorithm.
    We show that the resulting data releases are optimal when expected distortion quantifies utility, and locally optimal when mutual information quantifies utility.
    Numerical experiments on real data demonstrate that the proposed adaptive algorithm can exploit previously released information to reduce cumulative leakage under collusion without sacrificing much, if any utility.
    Finally, we discuss how our findings may extend to applications in machine learning.
\end{abstract}

\section{Introduction}
In data privacy literature, a common goal is to maximise the utility of a data release, whilst maintaining a certain level of privacy.
In its simplest form, the problem considers a single party that has made one request for data.
They are sent a version of their request that has undergone some transformation via a privacy mechanism.
Should this party make a second data request, the naive approach of using the same privacy mechanism poses clear issues.
Assuming the mechanism was designed subject to a privacy budget, we can expect that applying it twice will not meet the same budget. 
In fact, in the worst case, the two data releases may combine synergistically, revealing far more information than the sum of their individual contributions.\footnote{To see this, consider the general possibility that $I(X,Y;Z) > I(X;Z) + I(Y;Z)$ for random variables $X, Y, Z$.}
A malicious actor could design their data requests to exploit this fact.
To address this, many authors have considered adaptive privacy schemes, which take into account previous data releases.

It is also conceivable that multiple distinct parties may be interested in a dataset. 
For example, a medical dataset might prove useful to several research groups, each pursuing different objectives.
The data handler may wish to exploit the full research potential of the dataset rather than limiting access to one group.
Likewise, social media companies sell or license users' data to multiple third parties for advertising and analytics purposes.
In both cases, data handlers are unlikely to communicate with all parties simultaneously and will instead receive their data requests sequentially.
Recognising that each party, such as a research group or third party, is a potential adversary, we will use the terms \textit{party} and \textit{adversary} according to the context.
The data handler should consider the possibility of data sharing (collusion) amongst the adversaries, which would undermine individual privacy mechanisms.
In the medical dataset example,
a research student could inadvertently share their dataset with a collaborating group, who could combine it with their own information, revealing more than was authorized.
It is also conceivable that a student could be part of two groups.
In the social media example,
third parties may combine their social media data in an unauthorised manner to gain a competitive advantage,
or a breach affecting multiple parties could allow a single attacker to access several datasets.
Although measures should be taken to prevent such instances of collusion, some risk will remain.
Some level of protection should still be afforded to individuals' data in the worst case.
To inform collusion constraints on privacy mechanisms, the data handler may consider the likelihood of collusion, as well as their maximum tolerance for information leakage in the worst case that all adversaries combine their data.

\begin{figure}[t]
  \centering
\includegraphics[scale=0.5]{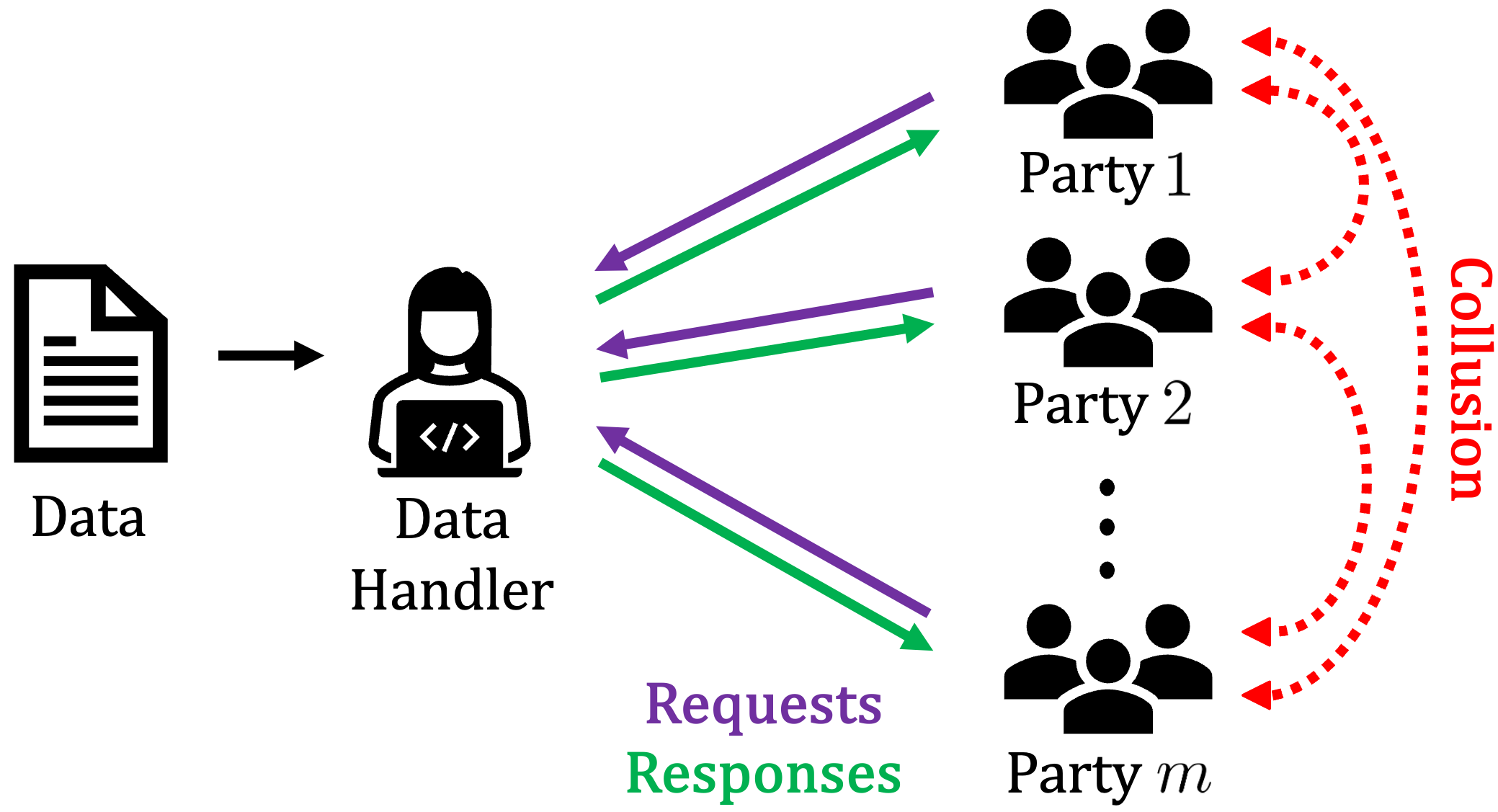}
  \caption{Problem setup for a multi-party adaptive scheme}
  \label{fig: simple setup}
\end{figure}

We consider a scenario in which a data handler wishes to provide useful information to $m$ parties, whilst preserving privacy and guarding against collusion.
As illustrated in Figure \ref{fig: simple setup},
$m$ distinct parties make requests on the data sequentially.
This constitutes an online scenario, as future requests are not known at the time of a data release.
The data handler must not compromise the privacy of the database, but would like to maximise the utility of the data released to each party,
where both expected distortion and mutual information are considered as utility metrics.
We adopt an attack model under which one malicious actor \textit{may} gain access to the data supplied to all parties up to the most recent release.
This is a worst case model and does not assume any knowledge of which parties are likely to collude. Other variations of the collusion attack are also possible and are discussed in Section \ref{section: collusion}.
% as a direction for further work.
For each data release, the privacy leakage, measured by mutual information, to the intended recipient is limited to some value.
To guard against collusion, the total information an adversary could obtain from all releases to date is bounded as a worst-case measure.
The method is adaptable, and constraints may be adjusted by the data handler based on their requirements.

\subsection{Prior Work}
Seminal work on the privacy-utility trade-off is presented by Yamamoto in \cite{Yamamoto}.
The paper considers a database with private and useful information. It proposes an encoder–decoder problem seeking to balance minimal distortion of the useful data against maximal uncertainty of private data, measured with conditional entropy (equivocation).
As such, it casts the problem as a rate-distortion-equivocation problem.
Later, Sankar et al. \cite{Sankar} generalised the framework, for instance allowing for side information.
Both papers sketch out a feasible privacy-utility trade-off region.
More recent works have refined the characterisation of privacy--utility trade-offs using information-theoretic, learning-based, and multi-agent formulations \cite{Kalantari2018,Yang2024,Wang2021}.
These works differ from ours in that they consider one-shot disclosure to a single recipient.
Thus, the problem is not sequential.

Several studies \cite{Dwork2010,Riboni2012,Erdogdu2015,Hasan2017,Dong2024,SHMUELI2015, LocationTrace} have explored sequential privacy mechanisms to protect against a single potential adversary or interested party. An early contribution in this area was by Dwork et al. \cite{Dwork2010}, who examined changing data, such as traffic conditions, under continual observation. They developed an algorithm that guarantees differential privacy for counting queries in these settings.
This work inspired numerous studies on continual-observation differential privacy, including the recent work of Dong et al. \cite{Dong2024}, which considered more complex join queries, where join queries combine data from multiple tables through related attributes.
In \cite{LocationTrace}, location trace data is studied, which is naturally sequential.
Their framework parallels ours in using distortion and mutual information to quantify utility and privacy.
These studies, as well as \cite{Riboni2012,Erdogdu2015,Hasan2017}, are motivated by time-changing data.
Conversely, Shmueli et al. \cite{SHMUELI2015} study sequential requests on a static dataset, as we do. They develop an algorithm that generates responses preserving $\ell$-diversity.
Whilst their approach supports sequential releases, their anonymisation is not adaptive; each output does not depend on previous ones.
Instead, a fixed set of  $\ell$-diverse tables with the same quasi-identifiers and sensitive values is precomputed. Each subsequent data release must be consistent with this set.
In contrast, a focus of our work is to exploit adaptivity for improved utility subject to privacy constraints.
More broadly, the works discussed above do not account for data requests made by distinct parties.
As such, there can  be no notion of collusion.

Relatively few studies consider multiple interested parties making data requests.
In \cite{Zamani2021}, multiple potential adversaries request different parts of the same database. As revealed data is assumed to be public, only one response is produced and sent to all parties.
This is not true of \cite{Lu2018}, in which parties can sequentially query a database and receive their own noisy response. The noise parameter is chosen to satisfy differential privacy and before the next query, the privacy budget is updated.
Although this method is adaptive, each data release remains independent of the actual outcomes of previous releases.
Additionally, the privacy budget depletes identically whether a sequence of requests comes from a single party or many different parties.
Like \cite{Zamani2021}, it is essentially assumed that 
any revealed data becomes public. 
This is akin to assuming that all parties share their data with all other parties, which we call \emph{full collusion}.
In many ways, this problem reduces to that of a single party,
as all parties are essentially treated as a single entity with respect to privacy. 
In our work, we do not make the same assumption, which enables better utility.
We seek to limit the privacy leakage to each interested party, as well as regulating the possible effects of collusion.
To the best of our knowledge, this scenario has not yet been considered.

\subsection{Contributions}

Our main contributions are as follows.
\begin{enumerate}
    \item We propose a privacy-utility framework for sequential data requests by multiple parties, 
    guarding against the worst case in which a malicious actor gains access to all data disclosed up to the most recent release.
    The framework supports two notions of utility, distortion and mutual information, with justification provided for both, and is cast as an optimisation problem.
    \item 
    Within the framework, we formulate the dual problem and focus on solving its its inner minimisation.
    We adapt the well-known Blahut--Arimoto algorithm to find solutions under fixed Lagrange multipliers. The solutions are optimal when expected distortion is the utility metric and locally optimal when mutual information is the utility metric.
    We demonstrate the Blahut--Arimoto algorithm's effectiveness via numerical simulation.
    \item We propose a sequential data release algorithm that provides a systematic mechanism for responding to each data request in turn, while maintaining privacy guarantees.
    The algorithm applies Blahut--Arimoto-style iterations for a set of Lagrange multipliers, and uses a targeted bisection 
    search
    to map 
    the desired privacy constraints to the corresponding Lagrange multipliers.
    Under the expected distortion utility metric, the algorithm is optimal.
    \item We provide empirical validation for the sequential algorithm on real data, demonstrating that it can exploit previously released information to reduce cumulative leakage under collusion without sacrificing utility.
    Benefits of the procedure are highlighted through comparison with non-adaptive and symmetric-channel baselines.
\end{enumerate}
Finally, we discuss a potential application of our work in machine learning.
In particular, we highlight a connection between the dual inner minimisation and the structure of progressive neural networks.
This indicates the broader relevance of the dual inner minimisation beyond serving as a tool for solving the privacy problem.

\section{Problem Setup} \label{section: problem setup}
We use capital letters to represent random variables, and lowercase letters for their realisations. 
Curly brackets denote a probability distribution, for example $\{p(x)\}$ is a probability vector whilst $p(x)$ is the probability of the realisation $X=x$.
Script font, such as $X \in \mathcal{X}$, is used to denote alphabets.
Finally, superscripts indicate sequences, for example $x^k = (x_1, \dots, x_k)$. 

The problem setup, as illustrated in Figure~\ref{fig: simple setup}, consists of $m$ parties making sequential requests on a database $X$ with probability distribution $\{p(x)\}$. Party $1$ makes the first request, followed by party $2$, and so on.
As she does not have knowledge of future requests at the time of a data release, the data handler cannot jointly optimise all releases.
Instead, she must apply an adaptive privacy scheme, meaning
the data released to each party depends on the database, $X$, the data request, and all previous data releases.

\begin{figure}[t]
  \centering
\includegraphics[scale=0.55]{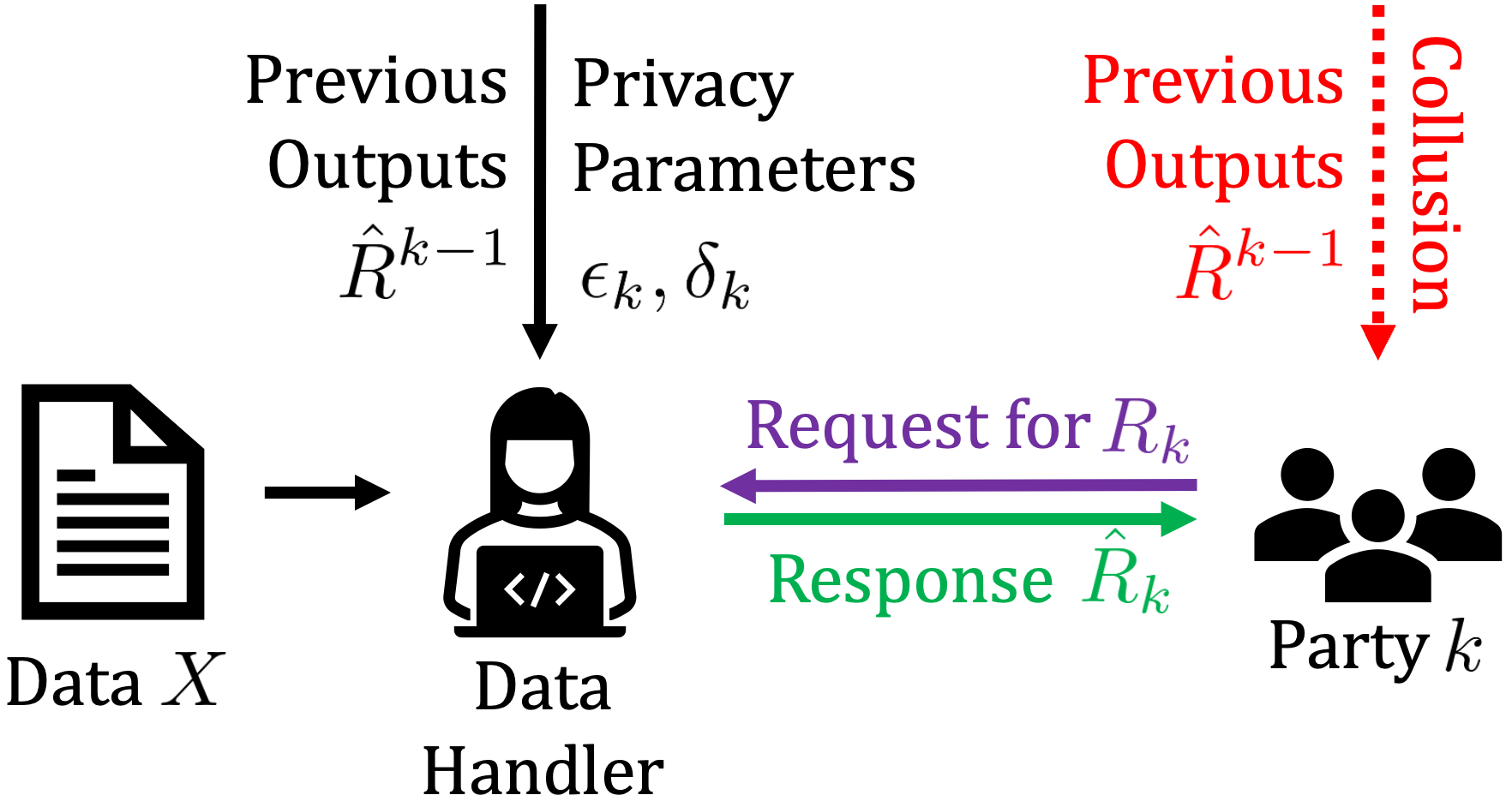}
  \caption{Multi-party adaptive privacy scheme: a detailed view of the $k$th party}
  \label{fig: setup}
\end{figure}

Figure \ref{fig: setup} provides a detailed view of the $k$th step of the adaptive scheme, where $k \in \{1, \dots, m\}$.
At the $k$th request, $R_k$, which is correlated with $X$, is the data value of interest to party $k$.
We do not assume that $R_k$ is a function of the database, and instead allow it to be arbitrarily correlated with $X$.
In a medical dataset, $R_k$ may be the proportion of patients aged over $65$ with a dementia diagnosis.
Alternatively, $R_k$ could be a subset of $X$, namely the age and diagnosis attributes.
In response to this, the data handler releases $\hat{R}_k$, taking into account the database $X$, previous outputs $\hat{R}^{k-1} = (\hat{R}_1, \dots, \hat{R}_{k-1})$, and privacy parameters $\epsilon_k$ and $\delta_k$. Let $\{p(r_k, \hat{r}^{k-1}, x)\}$ denote the joint distribution of these random variables. 

The data handler would like to maximise the utility to party $k$.
Let $U(\hat{R}_k, R_k)$ represent the utility of $\hat{R}_k$ to a party that requested $R_k$.
We will consider expected distortion and mutual information as means to quantify utility. Furthermore, the data handler must not compromise the privacy of the database.
We limit the individual leakage to party $k$, as well as the leakage to the first $k$ parties, should they all share their data.
The leakages, measured with mutual information, can be no larger than $\epsilon_k$ and $\delta_k$ respectively, where $\epsilon_k \leq \delta_k$.\footnote{Trivially, for the first request $\epsilon_1=\delta_1$ and both leakage constraints collapse into one.}
The information leakage to the current party is $I(\hat{R}_k; X)$.
We note that, in the worst case scenario, party $k$ could gain access to $\hat{R}^{k-1}$ through collusion.
In this case, the leakage is $I(\hat{R}_k, \hat{R}^{k-1}; X)$.

Upon receiving a request for $R_k$, the data handler solves
\begin{align}
\min_{\{p(\hat{r}_k \mid \hat{r}^{k-1}, x)\}} \quad & - U(\hat{R}_k, R_k) \nonumber \\
\text{s.t.} \quad & I(\hat{R}_k; X) \leq \epsilon_k \label{eq: optimisation problem} \\
& I(\hat{R}_k, \hat{R}^{k-1} ; X) \leq \delta_k, \nonumber
\end{align}
where $\{p(\hat{r}_k \mid \hat{r}^{k-1}, x)\}$ is the conditional probability distribution,
and the utility function takes one of two forms.
The first is expected distortion.
A distortion function $d$ is a non-negative function defined on $\hat{\mathcal{R}}_k \times \mathcal{R}_k$.
In this case the utility is
\begin{align*}
    U(\hat{R}_k, R_k) = - \mathbb{E} \left[ d(\hat{R}_k, R_k) \right].
\end{align*}
Clearly, a smaller distortion corresponds to better utility for the interested party.
Note that zero average distortion may not be attainable, as the data handler operates on the database $X$ while $R_k$ need not be a function of it.
An interested party may prioritise distortion as an objective because it enforces interpretability of the data. In contrast, measures that optimize only for correlation, such as mutual information, do not guarantee that the receiving party can effectively use or act on the data.
Alternatively, mutual information may be favoured as it preserves overall information content, promoting utility beyond a single notion of distortion.
Here, the utility is given by
\begin{align*}
    U(\hat{R}_k, R_k) = I (\hat{R}_k; R_k).
\end{align*}
Without explicit knowledge of the interested parties' goals, we examine both metrics separately.
We call the two resulting optimisation problems the \emph{expected distortion problem} and the \emph{mutual information problem} respectively.
It is worth noting that the former is a convex problem, whilst the latter is not.

Having solved the appropriate problem, the data handler releases $\hat{R}_k$ according to the optimal $\{p(\hat{r}_k \mid \hat{r}^{k-1}, x)\}$.
Constants $\epsilon_k$ and $\delta_k$ should be tuned based on the relative importance of utility, leakage, and worst case leakage.
For example, if privacy is the main priority, both constants should be small.
If collusion is deemed very unlikely, $\delta_k$ may be much larger than $\epsilon_k$.
In any case, it must be that $\epsilon_k \leq \delta_k$, and
the worst case leakage budget, $\delta_k$, must not decrease with increasing $k$.

In this paper, our objective is to determine the optimal strategy for each data release, namely the distribution $\{{p(\hat{r}_k \mid \hat{r}^{k-1}, x)}\}$ for each $k \in \{1,..., m \} $.
In doing so, we can characterize the trade-off by sketching either the distortion–privacy–collusion curve $\mathsf{D}_k(\epsilon_k, \delta_k)$ or the information–privacy–collusion curve $\mathsf{I}_k(\epsilon_k, \delta_k)$, where $\mathsf{D}_k$ denotes the minimum achievable expected distortion and $\mathsf{I}_k$ the maximum achievable mutual information.

\section{Guiding Principles for Data Release}
\label{section: procedure}

To gain an intuition for the optimal data release, let us first consider an example. 
Suppose a request for $R_k$ is received such that $R_k = R_1$. In other words, party $k$ makes the same request as party $1$.
Assume further that both parties are granted the same individual leakage budget, i.e., $\epsilon_k = \epsilon_1$.
In this scenario, the optimal data release is $\hat{R}_k = \hat{R}_1$.
To see this, consider the optimisation problem corresponding to party $k$ with $R_k=R_1$ and $\epsilon_k = \epsilon_1$:
\begin{align*}
\min_{\{p(\hat{r}_k \mid \hat{r}^{k-1}, x)\}} \quad & - U(\hat{R}_k, R_1) \\
\text{s.t.} \quad & I(\hat{R}_k; X) \leq \epsilon_1 \\
& I(\hat{R}_k, \hat{R}_1 ; X) \leq \delta_k,
\end{align*}
which imposes stricter constraints than that of party 1:
\begin{align*}
\min_{\{p(\hat{r}_1 \mid x)\}} \quad & - U(\hat{R}_1, R_1 ) \\
\text{s.t.} \quad & I(\hat{R}_1; X) \leq \epsilon_1.
\end{align*}
Conditioning on $\hat{r}^{k-1}$ in party $k$'s optimisation problem can help satisfy the additional constraint, but it cannot reduce the value of the objective function below the minimum achievable of party 1's optimisation. Consequently, party $k$'s optimisation problem cannot achieve a lower minimum than that of party 1's.
We can achieve the same minimum by setting $\hat{R}_k = \hat{R}_1$.  
This choice is feasible because it ensures $I(\hat{R}_k, \hat{R}^{k-1}; X) = I(\hat{R}^{k-1}; X) \leq \delta_{k-1} \leq \delta_k$, given that the previous release was selected to satisfy its constraints.

This example provides insight into the effect of the collusion constraint: it promotes the reuse of information across data releases whenever doing so helps to improve utility.
This reduces the adversaries' power of learning through collusion.

\subsection{General Optimisation Problem}
To facilitate discussions, we let $R$, $\hat{R}$, $\epsilon$ and $\delta$ represent the data request, data release and privacy parameters corresponding to an arbitrary transaction in a sequence of $m$ such transactions. Similarly, we use $Z$ to collectively denote all previous data releases.
Let $\{p(r, z, x)\}$ be the joint distribution of $R$, $Z$ and the database $X$.
The optimisation problem \eqref{eq: optimisation problem} becomes
\begin{align}
\min_{\{p(\hat{r} \mid z, x)\}} \quad & - U(\hat{R}, R) \nonumber \\
\text{s.t.} \quad & I(\hat{R}; X) \leq \epsilon \label{eq: optimisation problem new vars} \\
& I(\hat{R}, Z ; X) \leq \delta. \nonumber
\end{align}
This formulation represents the problem solved at each sequential step, and
solving it will give us a procedure for the adaptive problem.

When expressed as in \eqref{eq: optimisation problem new vars}, the expected distortion problem shares obvious similarities with the rate-distortion problem \cite{Blahut72} (or, more accurately, its inverse formulation, the distortion-rate problem).
Without the second constraint, the conditioning on $z$ becomes irrelevant and problem \eqref{eq: optimisation problem new vars} with $U(\hat{R}, R)=-E[d(\hat{R}, R)]$ is equivalent to the distortion-rate problem.  With the interpretation that $\hat{R}$ is a compressed version of $X$, \eqref{eq: optimisation problem new vars} can be thought of as the minimization of the average distortion between $\hat{R}$ and $R$, where $\hat{R}-X-R$ forms a Markov chain, while limiting the compression rate $I(\hat{R}; X)$.

Consider the Lagrange dual problem of \eqref{eq: optimisation problem new vars}:
\begin{align}\label{eq: opt prob}
    \max_{\mu_1, \mu_2 \geq 0} \; \min_{ \{p(\hat{r}\mid z,x)\} }  \Big(- U (\hat{R}&, R) + \mu_1 (I(\hat{R} ; X) - \epsilon) \nonumber \\&+ \mu_2 (I(\hat{R}, Z ; X) - \delta) \Big),
\end{align}
where $\mu_1$ and $\mu_2$ are Lagrange multipliers. We refer to the inner objective function as $g(\{p(\hat{r} \mid  x, z)\})$.
We can draw parallels between the inner minimisation in \eqref{eq: opt prob} and the information bottleneck (IB) method \cite{tishby2000}, particularly when $U(\hat{R}, R) = I(\hat{R};R)$.
We can see this more clearly if we temporarily omit constant terms leaving
\begin{align*}
    \min_{p(\hat{r} \mid  z,x)} \! \left( - I(\hat{R}; R) + \mu_1 I(\hat{R} ; X) + \mu_2 I(\hat{R}, Z ; X) \right).
\end{align*}
By setting $\mu_2=0$, we can see that this is equivalent to the IB problem.
With the same interpretation that $\hat{R}$ is a compressed version of $X$ such that $\hat{R} - X -R$ forms a Markov chain, the IB problem aims to maximize the information $I(\hat{R}; R)$ about $\hat{R}$ in $R$ whilst encouraging compression by penalising $I(\hat{R}; X)$.
%In the IB problem, $\hat{R}$ is chosen according to the Markov chain $\hat{R} - X - R$ so as to compress $X$ whilst retaining information about $R$.
%An optimisation is carried out to minimise the function: $I(\hat{R};X) - \beta I(\hat{R}; R)$, where $\beta$ is some constant.
%In our problem, we see a similar Markov chain: $\hat{R} - (X, Z) - R$, where in both cases $R$ is arbitrarily correlated with $X$. 
%In contrast with the IB problem, in limiting $I(\hat{R}; X)$, we only desire compression of $X$, and are not directly interested in that of $Z$. 
%Also, we seek to minimise the distortion between $\hat{r}$ and $r$ rather than to maximise their mutual information.
%As mentioned in Section \ref{section: mathematical framework},
%we made this decision for the sake of the usability of the data for the requesting party. Even if $\hat{R}$ and $R$ share substantial mutual information, this does not necessarily translate into a clear understanding of specific values of $r$.
%As previously mentioned, whilst mutual information quantifies statistical dependence, it does not provide a decoding rule or indicate how to extract knowledge of actual outcomes.
% should we still say this ^ ?
%As a result, the IB problem is largely applied in machine learning, where information shared can be prioritised over interpretability.
%The $I(\hat{R}, Z;X)$ term is distinct from both existing problems. 

Our sequential optimisation problem, though motivated by a privacy setting, can also be viewed as a generalization of the rate-distortion or IB problems. We see a similar Markov chain: $\hat{R}-(X, Z)-R$, where $\hat{R}$ depends also on all the previous data releases $Z$ instead of just $X$. Additionally, we are concerned with the overall compression rate for all data releases $(\hat{R}, Z)$ through the second constraint.

\subsection{Solution Overview}
\label{section: solution overview}
To solve our optimisation problem \eqref{eq: optimisation problem}, we will take inspiration from methods developed to handle both the rate distortion and IB problems.
The general method, for example, to compute a rate-distortion curve \cite{Blahut72} involves the use of a Blahut-Arimoto (BA) style algorithm. This, in principle, allows us to identify all the points of the curve, where each point corresponds to a distinct value of the Lagrange multiplier.
In practice, a finite number of points are first identified using the BA-style algorithm and then an arbitrary point on the curve is located via bisection search.
We propose a similar procedure to solve our adaptive problem.

For each data request, we reformulate the problem to its current step formulation \eqref{eq: optimisation problem new vars}.
We then convert the problem to its dual, as in \eqref{eq: opt prob}.
For each Lagrange multiplier pair $(\mu_1, \mu_2)$ in a predefined grid, we use a modified Blahut-Arimoto-style algorithm (see Section \ref{section: BA algorithm}) to obtain a conditional data release distribution. 
From this, we can compute the corresponding utility and leakage values, which can be used to generate a coarse approximation to the distortion–privacy–collusion or information–privacy–collusion curve.
We can find the solution corresponding to the particular constraints using a bisection search.
It is worth remarking that in the case of the expected distortion measure, the curve obtained by solving the dual is precisely the distortion-privacy-collusion curve because the primal problem \eqref{eq: optimisation problem new vars} is convex. On the other, the primal problem is not convex with mutual information, so the curve obtained via this approach will bound the information-privacy-collusion curve from below.

The sequential data release algorithm is provided in Section \ref{section: sequential solution procedure}.
In the following section, we first derive the Blahut-Arimoto-style algorithm.

\section{Blahut-Arimoto-Style Algorithm} \label{section: BA algorithm}
The Blahut-Arimoto (BA) algorithm is an iterative algorithm proposed by Blahut \cite{Blahut72} to solve the rate-distortion problem and has since been generalised to deal with the IB problem \cite{tishby2000}.
Zhang et al. \cite{LocationTrace} make use of it to numerically solve their location trace privacy problem.
Here, we derive a BA-style iterative algorithm to solve the inner minimisation of the dual problem \eqref{eq: opt prob}.

First, we deal with the expected distortion problem:
\begin{align} \label{eq: inner min}
    \min_{\{p(\hat{r}\mid z,x)\}} g(\{p(&\hat{r}\mid z,x)\}) = \min_{ \{p(\hat{r}\mid z,x)\} }  \Big(\mathbb{E} \left[d(\hat{R}, R) \right] + \nonumber \\& \! \mu_1 (I(\hat{R} ; X) - \epsilon)+ \mu_2 (I(\hat{R}, Z ; X) - \delta) \Big).
\end{align}
We outline an implicit solution to the problem in the form of a self consistent equation.
By treating related variables as independent, it is possible to iteratively update each variable towards a solution.
We provide the necessary update equations and prove convergence to a consistent solution.
We remark that, as the expected distortion problem is convex, its BA-style algorithm will converge to the global minimum.
This is not the case for the mutual information problem, for which a local minimum may be found.

\subsection{An Implicit Solution}
\label{subsection: an implicit solution}
Although an explicit analytical solution to our dual problem cannot be found, we can provide an exact implicit solution. 
The solution is implicit in the sense that it is not a closed form expression for $p(\hat{r}  \mid  z, x)$; it is a self consistent equation. Let $p^\star(\hat{r} \mid  z, x)$ be the optimal assignment in \eqref{eq: inner min}.

\newtheorem{proposition}{Proposition} 
\begin{proposition} \label{th: formal solution}
    The following is a sufficient condition on $p^\star(\hat{r} \mid  x, z)$:
    \begin{align} \label{eq: formal solution}
        p^\star(\hat{r} \mid  z,x) = \frac{1}{\eta} \Big[ p^\star(\hat{r})^{\mu_1}& p^\star(z \mid  \hat{r}, x)^{\mu_1} p^\star(\hat{r} \mid z)^{\mu_2} \nonumber \\& 2^{-\sum_r p (r \mid z, x) d(\hat{r}, r)} \Big]^{\frac{1}{\mu_1 + \mu_2}},
    \end{align}
    where $\eta = \eta (x, z, \mu_1, \mu_2)$ is a normalisation factor:
    \begin{align*}
        \eta = \sum_{\hat{r}} \Big[  p^\star(\hat{r})^{\mu_1} p^\star(z \mid  \hat{r}&, x)^{\mu_1} p^\star(\hat{r} \mid z)^{\mu_2} \nonumber \\& 2^{-\sum_r p (r \mid z, x) d(\hat{r}, r)} \Big]^{\frac{1}{\mu_1 + \mu_2}}.
    \end{align*}
\end{proposition}
The condition is obtained by taking the derivative of the function to be minimised and setting it equal to zero.
As the function $g(\{p(\hat{r}\mid z,x)\})$ to be minimised is convex, the condition is sufficient.
Let
\begin{align*}
    \mathcal{L}(&p(\hat{r}  \mid  z,x), \bm{\lambda}) = \mathbb{E} \left[ d(\hat{R}, R) \right] + \mu_1 I(\hat{R};X) \; + \\& \mu_2 I(\hat{R}, Z; X) + \sum_{z,x} \lambda(z, x) \left( \sum_{\hat{r}} p(\hat{r} \mid z,x) - 1 \right),
\end{align*}
where the final term has been added to constrain $p(\hat{r} \mid  x, z)$ to valid probability distributions.
We require the derivatives of $\mathbb{E} [d(\hat{R}, R)]$, $I(\hat{R};X)$ and $I(\hat{R}, Z; X)$ with respect to $p(\hat{r}  \mid  z,x)$. 
Starting with $\mathbb{E} \big[ d(\hat{R}, R) \big]$:
\begin{align*}
    \mathbb{E} \left[ d(\hat{R}, R) \right] = \sum_{\hat{r}, z, x, r} p(\hat{r}  \mid  z, x) p(r, z, x) d(\hat{r}, r),
\end{align*}
where we have used the fact that $\hat{R}- (Z, X) - R$ is a Markov chain. We obtain
\begin{align*}
    \frac{\partial}{\partial p(\hat{r}  \mid  z, x)} \mathbb{E} \left[ d(\hat{R}, R) \right] \! = p(z, x) \sum_r p(r \mid z, x) d(\hat{r}, r).
\end{align*}
Next, 
\begin{align*}
    &I(\hat{R};X) =\\& \sum_{\hat{r}, z, x} p(\hat{r} \mid z,x)p(z,x) \log \frac{\sum_{\hat{z}} p(\hat{r}  \mid  \hat{z}, x) p(\hat{z} | {x})}{\sum_{x',z'} p(\hat{r}  \mid  z', x') p(z', x')}.
\end{align*}
Taking the derivative gives
\begin{align*}
    \frac{\partial}{\partial p(\hat{r}  \mid  z, x)} I(\hat{R};X) = p(z,x) \log \frac{p(\hat{r} \mid x)}{p(\hat{r})}.
\end{align*}
Lastly, we find the derivative of $I( \hat{R}, Z; X)$. Using
\begin{align*}
    I(\hat{R}&, Z; X) = \\& \sum_{\hat{r}, z, x} p (\hat{r} \mid  z, x) p(z, x) \log \frac{p (\hat{r} \mid  z, x) p(z  \mid x)}{\sum_{x'} p(\hat{r}  \mid z, x') p(z, x')}
\end{align*}
yields
\begin{align*}
    \frac{\partial}{\partial p(\hat{r}  \mid  z, x)} I(\hat{R}, Z ;X) = p(z,x) \log \frac{p( \hat{r}, z  \mid x)}{p(\hat{r}, z)}.
\end{align*}
We can finally return to $\mathcal{L}(p(\hat{r}  \mid  z,x), \bm{\lambda})$. Setting the derivative to zero gives
\begin{align*}
    p (z,x) \bigg\{ \sum_r p(r \mid z, & \, x) d(\hat{r}, r) + \mu_1 \log \frac{p(\hat{r} \mid x)}{p(\hat{r})} \;+ \\& \mu_2 \log \frac{p(\hat{r}, z  \mid x)}{p( \hat{r}, z)} + \frac{\lambda(z, x)}{p(z, x)} \bigg\} = 0,
\end{align*}
or equivalently
\begin{align*}
    \sum_r p(r \mid z, x) d(\hat{r}, r) +\log &\frac{p(x \mid  \hat{r})^{\mu_1} p(\hat{r} \mid  z, x)^{\mu_2}}{ p(\hat{r} \mid  z)^{\mu_2}} \\& \qquad\qquad + \lambda'(z, x) = 0,
\end{align*}
where functions of $z$ and $x$ have been absorbed into $\lambda'$.
To match Proposition \ref{th: formal solution}, we need to express this in terms of $p(\hat{r} \mid  z, x)$, $p(\hat{r} \mid z)$, $p(\hat{r})$, and $p(z \mid \hat{r}, x)$. 
Using the identities
\begin{align*}
    &p(x \mid \hat{r}) = \frac{p(\hat{r} \mid x) p(x)}{p(\hat{r})},
    &\frac{p(\hat{r} \mid x)}{p(z \mid x)} = \frac{p(\hat{r} \mid  z, x)}{p(z \mid \hat{r}, x)},
\end{align*}
in turn straightforwardly gives
\begin{align*}
    \sum_r p(r \mid z, x) d(\hat{r}, r) + \log& \frac{p(\hat{r} \mid  z, x)^{\mu_1 + \mu_2}}{p(\hat{r})^{\mu_1} p(z \mid  \hat{r}, x)^{\mu_1} p(\hat{r}  \mid  z)^{\mu_2}} \\& \qquad\qquad\;\: + \lambda''(z, x) = 0,
\end{align*}
where functions of $z$ and $x$ have been absorbed into $\lambda''$.
Rearranging, and choosing $\lambda''$ such that $\sum_{\hat{r}} p(\hat{r} \mid  z, x) = 1$ yields Proposition \ref{th: formal solution}.

\subsection{Self Consistent Variables}

In Proposition \ref{th: formal solution}, we have a self consistent equation where terms on the right hand side can be expressed in terms of the left hand side, $p(\hat{r} \mid  z, x)$.
More precisely,
\begin{align} \label{eq: self consistent 1}
    p(\hat{r}) & = \sum_{z, x} p(\hat{r} \mid  z, x) p(z, x), \\
    p(z  \mid  \hat{r}, x) &= \frac{p(\hat{r} \mid  z, x) p(z \mid x)}{\sum_{z'} p(\hat{r} \mid  z', x) p(z' \mid x)}, \\
    p(\hat{r} \mid  z) & = \sum_x p(\hat{r} \mid  z, x) p(x  \mid  z).
    \label{eq: self consistent 4}
\end{align}
We will show that an iterative algorithm which treats $p(\hat{r} \mid  z, x)$ and each of the above terms as independent variables and updates them in turn will converge to a solution. Further, we will show that the objective function decreases with each update. 
First, we must express our objective function in terms of the variables given above.

\newtheorem{lemma}{Lemma}
\begin{lemma}\label{lemma: rearrange objective}
    The objective function, $g(\{p(\hat{r} \mid  z, x)\})$, in \eqref{eq: inner min} can be equivalently expressed as
        \begin{align}\label{eq:def:func:f}
        &\sum_{\hat{r}, z, x} p(\hat{r}, z, x) \log\frac{2^{\sum_r p(r \mid z, x) d(\hat{r}, r)} p(\hat{r} \mid  z, x)^{\mu_1 + \mu_2}}{p(z |  \hat{r}, x)^{\mu_1} p(\hat{r})^{\mu_1}  p(\hat{r} | z)^{\mu_2}} \! + \theta \nonumber \\
        & \triangleq f \big( \{p(\hat{r} |  z, x)\}; \{p(\hat{r})\}; \{p(z |  \hat{r}, x)\}; \{p(\hat{r} | z)\} \big)
    \end{align}
    where $\theta = \theta(\mu_1, \mu_2, \{p(z,x)\})$ is not a function of the distributions $\{p(\hat{r} |  z, x)\}$, $\{p(\hat{r})\}$, $\{p(z |  \hat{r}, x)\}$, and $\{p(\hat{r} | z)\}$.
    \begin{proof}
    Starting with non-constant terms in \eqref{eq: opt prob} and rearranging gives
        \begin{align*}
            &\mathbb{E} \left[ d(\hat{R}, R) \right] + \mu_1 I(\hat{R} ; X) + \mu_2 I( \hat{R}, Z ; X) \\
            &= \mathbb{E} \left[ d(\hat{R}, R) \right] + (\mu_1 + \mu_2) I( \hat{R}, Z ; X) - \mu_1 I(Z; X |  \hat{R})
        \end{align*}
        \begin{samepage}
        \begin{align*}
            &= \! \mathbb{E} \! \bigg[\! \sum_r p(r | z, x) d(\hat{r}, r)\! + \!(\mu_1 \!+\! \mu_2) \log \frac{p(\hat{r} |  z, x) p(z, x)}{p(\hat{r}  |  z) p(z) p(x)} \\& \qquad\qquad\qquad\qquad\qquad\qquad\: - \mu_1 \log \frac{p(z  \mid  \hat{r}, x) p(\hat{r})}{p(\hat{r} \mid  z) p(z)} \bigg],
        \end{align*}
        \end{samepage}
        with slight abuse of notation in the final line,
        where $\mathbb{E}$ represents an expectation over $\{p(\hat{r}, x, z) \}$, from which the result follows.
    \end{proof}
\end{lemma}

\subsection{Independent Minimisations}

The algorithm depends on treating the variables of $f$ defined in \eqref{eq:def:func:f} as independent. 
Theorem \ref{th: ind min} provides the basis of this. 
When referencing independent variables, we use $q$s in the place of $p$s to denote the 2nd 3rd and 4th inputs to $f$.
The independent variables are $\{p(\hat{r} \mid  z, x)\}$, $\{q_1(\hat{r})\}$, $\{q_2(z \mid  \hat{r}, x)\}$, and $\{q_3(\hat{r} \mid z)\}$. 
For brevity, we refer to them as $p$, $q_1$, $q_2$, and $q_3$, respectively. 
Subscripts are used to emphasise that the probability distributions $p, q_1, q_2, q_3$ are not related to each other.

\newtheorem{theorem}{Theorem} 
\begin{theorem} \label{th: ind min}
    The minimisation in \eqref{eq: inner min} over $\{p(\hat{r} \mid  z, x)\}$ is equivalent to a series of independent minimisations over $\{p(\hat{r} \mid  z, x)\}$, $\{q_1(\hat{r})\}$, $\{q_2(z \mid  \hat{r}, x)\}$, and $\{q_3(\hat{r} \mid z)\}$ i.e.,
    \begin{align*}
        &\min_{p} g(p) = \min_{p} \min_{q_1} \min_{q_2} \min_{q_3} f(p; q_1; q_2;q_3 ),
    \end{align*}
    where, for any $p$, the minimising $q$ assignments are
    \begin{align*}
        q_1(\hat{r}) & = \sum_{z, x} p(\hat{r} \mid  z, x) p(z, x), \\
        q_2(z  \mid  \hat{r}, x) &= \frac{p(\hat{r} \mid  z, x) p(z \mid x)}{\sum_{z'} p(\hat{r} \mid  z', x) p(z' \mid x)}, \\
        q_3(\hat{r} \mid  z) & = \sum_x p(\hat{r} \mid  z, x) p(x  \mid  z).
\end{align*}
\end{theorem}

To prove Theorem \ref{th: ind min}, we require the following lemma, which will help us to identify the minimising distributions.

\begin{lemma} \label{lemma: minimiser}
    Let $A$ and $B$ be random variables that are subsets of $\{\hat{R}, Z, X\}$, i.e., $A, B \subseteq \{\hat{R}, Z, X\}$, where $B$ can be empty.
    Then,
    \begin{align*}
        \min_{q(a \mid b)} \sum_{\hat{r}, z, x} \! p(\hat{r}, z, x)  \log\!\frac{1}{q(a | b)} = \!  \sum_{\hat{r}, z, x} \! p(\hat{r}, z, x)  \log\!\frac{1}{p(a | b)},
    \end{align*}
    where the minimisation is over all conditional distributions $q(a \mid b)$ and $p(a \mid b)$ signifies the true distribution of $A$ given $B$.
    \begin{proof} We can write
        \begin{align} 
            \sum_{\hat{r}, z, x} p(\hat{r}, z, x) &\log\frac{1}{q(a \mid b)} - \sum_{\hat{r}, z, x} p(\hat{r}, z, x)  \log\frac{1}{p(a \mid b)} \nonumber \\
            &= \sum_{\hat{r}, z, x} p(\hat{r}, z, x)  \log\frac{p(a \mid b)}{q(a \mid b)} \nonumber \\
            &= \sum_{a, b} p(a, b)  \log\frac{p(a \mid b)}{q(a \mid b)} \label{eq: change exp variables} \\ 
            &= \mathbb{E}_B \left[ D_{KL}(p(a \mid b)  \mid  \mid  q(a \mid b) ) \right] \geq 0, \nonumber
        \end{align}
        where in \eqref{eq: change exp variables} we realise that any part of $\{\hat{R}, Z, X\}$ not in $A$ and $B$ will be eliminated through a summation. The final inequality is met with equality if and only if $p =q$.
        This proof is similar to \cite[Lemma 10.8.1]{Cover2006}.
    \end{proof}
\end{lemma}
We can now combine Lemmas \ref{lemma: rearrange objective} and \ref{lemma: minimiser} to prove Theorem \ref{th: ind min}. We can write
\begin{align} 
    &\min_p g(p) \nonumber 
    \\ &= \min_p \sum_{\hat{r}, z, x} p(\hat{r}, z, x) \log \Bigg[ \frac{2^{\sum_r p(r \mid z, x) d(\hat{r}, r)} }{p(z \mid  \hat{r}, x)^{\mu_1} p(\hat{r})^{\mu_1} }\times \nonumber \\ & 
    \qquad\qquad\qquad\qquad\qquad\quad \frac{p(\hat{r} \mid  z, x)^{\mu_1 + \mu_2}}{p(\hat{r} \mid z)^{\mu_2}}
    \Bigg] + \theta \label{eq: apply lemma rearrange}\\
    &= \min_p \sum_{\hat{r}, z, x} p(\hat{r}, z, x) \bigg[ \mu_1 \log \frac{1}{p(\hat{r})} + \mu_1 \log \frac{1}{p(z | \hat{r}, x)} \nonumber \\& \qquad\qquad\; +  \mu_2 \log\frac{1}{p( \hat{r} \mid z)} + (\mu_1 + \mu_2) \log p(\hat{r} \mid  z, x) \nonumber \\&  
    \qquad\qquad\qquad\qquad\qquad\qquad + 2^{\sum_r p(r \mid z, x) d(\hat{r}, r)}  \bigg] + \theta \nonumber \\
    &= \min_p \min_{q_1} \min_{q_2} \min_{q_3}  \sum_{\hat{r}, z, x} p(\hat{r}, z, x) \bigg[ \mu_1 \log \frac{1}{q_1(\hat{r})} + \nonumber \\&\quad \mu_1  \log \frac{1}{q_2(z |  \hat{r}, x)} +  \mu_2 \log\frac{1}{q_3( \hat{r}| z)} + 2^{\sum_r p(r \mid z, x) d(\hat{r}, r)} \nonumber 
    \\& \qquad\qquad\qquad\; + (\mu_1 + \mu_2) \log p(\hat{r} \mid  z, x) \bigg] + \theta  \label{eq: apply lemma min} \\
    &= \min_p \min_{q_1} \min_{q_2} \min_{q_3} f\left(p ; q_1; q_2;  q_3 \right), \nonumber
\end{align}
where $\theta = \theta \big( \mu_1, \mu_2, \{ p(z,x) \} \big)$, and \eqref{eq: apply lemma rearrange} and \eqref{eq: apply lemma min} apply Lemmas \ref{lemma: rearrange objective} and \ref{lemma: minimiser} respectively.

\subsection{An Iterative Algorithm} \label{subsection: iterative algorithm}
In our proposed BA-style algorithm, Algorithm \ref{alg: BA distortion}, each iteration consists of four update steps.
\begin{algorithm}[htpb]
\label{alg: BA distortion}
\caption{Blahut--Arimoto: Expected Distortion Problem}
\SetAlgoLined
\DontPrintSemicolon
Input: $\{p(r, z, x)\}$, $\mu_1, \mu_2 \geq 0$

Initialize $\{q_1^0(\hat{r})\}$, $\{q_2^0(z|\hat{r},x)\}$, $\{q_3^0(\hat{r}|z)\}$

\Repeat{convergence of $\{p(\hat{r}| z, x) \}$ }{

\For{\textnormal{\textbf{each}} $z, x, \hat{r}$}{
    \parbox[t]{\dimexpr\linewidth-2em}{%
    $\displaystyle
    \eta_{t} \gets \sum_{\hat{r}'} \Big[  q_1^{t}(\hat{r}')^{\mu_1} q_2^{t}(z \mid  \hat{r}', x)^{\mu_1} \\ \hspace*{3em} q_3^{t}(\hat{r}' \mid z)^{\mu_2} 2^{-\sum_r p (r \mid z, x) d(\hat{r}', r)} \Big]^{\frac{1}{\mu_1 + \mu_2}}
    $%
  }

  \parbox[t]{\dimexpr\linewidth-2em}{%
    $\displaystyle
    p^t(\hat{r}|z,x) \gets \frac{1}{\eta_t} \Big[
      q_1^t(\hat{r})^{\mu_1} q_2^t(z|\hat{r},x)^{\mu_1} q_3^t(\hat{r}|z)^{\mu_2} \\
      \hspace*{8em} 2^{-\sum_r p(r|z,x) d(\hat{r}, r)}
    \Big]^{\frac{1}{\mu_1 + \mu_2}}
    $%
  }

  \parbox[t]{\dimexpr\linewidth-2em}{%
    $\displaystyle
    q_1^{t+1}(\hat{r}) \gets \sum_{z',x'} p^t(\hat{r}|z',x') p(z',x')
    $%
  }

  \parbox[t]{\dimexpr\linewidth-2em}{%
    $\displaystyle
    q_2^{t+1}(z|\hat{r},x) \gets 
    \frac{p^t(\hat{r}|z,x) p(z|x)}
         {\sum_{z'} p^t(\hat{r}|z',x) p(z'|x)}
    $%
  }
  
\vspace{2mm}
  \parbox[t]{\dimexpr\linewidth-2em}{%
    $\displaystyle
    q_3^{t+1}(\hat{r}|z) \gets \sum_{x'} p^t(\hat{r}|z,x') p(x'|z)
    $%
    }

  }
}
Output: $\{p^T(\hat{r}| z, x) \}$
\end{algorithm}
We use $t$ to denote the update counter, with $T$ representing the number of iterations until convergence, and $\eta_t = \eta_t (x,z,\mu_1,\mu_2)$ is the normalisation factor.

To reason that the algorithm works, we make three remarks. For brevity, let $f_q = f(p; q_1; q_2; q_3)$, and $f_q^t = f(p^t;q_1^t; q_2^t; q_3^t)$.
We assume throughout that all variables are initially positive, i.e., all elements of $q_1^0, q_2^0$ and $q_3^0$ are greater than zero.
As a consequence of this and the nature of the update steps, we will only arrive at distributions for which all elements of $p^t, q_1^t, q_2^t$ and $q_3^t$ are greater than zero.
\newtheorem{remark}{Remark}
\begin{remark} \label{remark: fq decreases}
    Each step of the algorithm causes $f_q^t$ to decrease, unless it has already converged.
\end{remark}
\noindent In Theorem \ref{th: ind min}, we have already shown that steps 7, 8 and 9 in Algorithm \ref{alg: BA distortion} minimise $f_q$ with respect to $q_1$, $q_2$ and $q_3$ respectively.
It is easy to see that $f_q$ is a convex function in $p$ for fixed $q_1, q_2, q_3$.
Taking the derivative of $f_q$ with respect to $p$ and setting it to zero gives the equation in step 6 of Algorithm \ref{alg: BA distortion}.
Therefore, $f^t_q \leq f^{t-1}_q$ for at each time step $t$.
\begin{remark} \label{remark: bounded}
    Both $f_q$ and $g(p)$ are bounded from below.
    \begin{proof}
        By Theorem \ref{th: ind min}, the lower bound of $g \big( \{ p(\hat{r}  \mid  z,x ) \} \big)$ also serves as a lower bound on $f_q$.
Examining \eqref{eq: inner min}, we see that the former is straightforwardly given by $- (\mu_1 \epsilon + \mu_2 \delta ) $.
    \end{proof}
\end{remark}
\begin{remark} \label{remark: step 4 consistent}
    After each iteration $t$ of Algorithm \ref{alg: BA distortion}, the variables $p^t, q_1^t, q_2^t$ and $q_3^t$ are consistent.
    Therefore, it follows from the definitions of the functions that $g(p^t)=f_q^t$.
\end{remark}
To argue that the algorithm converges to the optimiser $p^\star$ of $\min_p g(p)$, we consider two possibilities.
The first is that $g$ has its minimum at an internal point of its domain, i.e., $p^\star$ contains only positive entries.
Convergence is confirmed by Remarks \ref{remark: fq decreases} and \ref{remark: bounded}.
The limit satisfies the self-consistency equation \eqref{eq: formal solution}, which is a sufficient condition for the optimality of $\min_p g(p)$.
Thus, the algorithm converges to $p^\star$.

The second possibility is that the minimum of $g$ lies on the boundary of its domain, i.e., $p^\star$ contains at least one zero element.
In this case, the algorithm may not reach a point that satisfies the sufficient condition in Proposition \ref{th: formal solution}.
However, we still argue convergence to a distribution $p$ for which $g(p)$ is asymptotically close to $g(p^\star)$.
From Remark \ref{remark: fq decreases}, we know that eventually, $f_q^t$ must be asymptotically close to the minimum of $f_q$, which has the same value as the minimum of $g$.
Also, by Remark \ref{remark: step 4 consistent}, after each iteration $t$, the algorithm generates a consistent set of distributions $p^t$, $q_1^t$, $q_2^t$ and $q_3^t$, of which $p^t$ can be fed into the function $g$.
Therefore, eventually $f_q^t$ is both asymptotically close to the minimum of $f_q$ and equal to $g(p^t)$.
Hence, the output $p^T$ of the algorithm after running it for sufficiently large number of iterations $T$, will attain a value $g(p^T)$ that is arbitrarily close to the global minimum $g(p^\star)$.

\subsection{Blahut-Arimoto for the Mutual Information Problem}

To solve the Mutual Information problem, a data handler can employ Algorithm \ref{alg: MI problem} as the Blahut--Arimoto-style step.

\begin{algorithm}[!htpb]
\caption{Blahut--Arimoto for Mutual Information with Multiple Initialisations}
\label{alg: MI problem}
\SetAlgoLined
\DontPrintSemicolon

Input: $\{p(r, z, x)\}$, $\mu_1, \mu_2 \geq 0$ \;

\For{$i = 1, \dots, N_{\text{j}}$}{

    Initialize $\{q_1^0(\hat{r})\}$, $\{q_2^0(z|\hat{r},x)\}$, $\{q_3^0(\hat{r}|z)\}$, $\{q_4^0(r|\hat{r})\}$ \;

    \Repeat{convergence of $\{p(\hat{r}| z, x) \}$ }{

      \For{\textnormal{\textbf{each}} $z, x, \hat{r}$}{

       \parbox[t]{\dimexpr\linewidth-2em}{%
        $\displaystyle
        \eta_t \gets \sum_{\hat{r}'} \Big[
          q_1^t(\hat{r}')^{\mu_1}
          q_2^t(z|\hat{r}',x)^{\mu_1}
         q_3^t(\hat{r}'|z)^{\mu_2}
          \\ \hspace*{4em} 2^{-
            D(
              \{p(r|z,x)\} ||
              \{q_4^t(r|\hat{r}')\}
            )
          }
        \Big]\! ^{\tfrac{1}{\mu_1+\mu_2}}
        $%
        }

        \parbox[t]{\dimexpr\linewidth-2em}{%
        $\displaystyle
        p^t(\hat{r}|z,x) \gets \frac{1}{\eta_t} \Big[
          q_1^t(\hat{r})^{\mu_1}
          q_2^t(z|\hat{r},x)^{\mu_1}
          \\ \hspace*{0.5em} q_3^t(\hat{r}|z)^{\mu_2}
          2^{-
            D(
              \{p(r|z,x)\} ||
              \{q_4^t(r|\hat{r})\}
            )
          }
        \Big]\! ^{\tfrac{1}{\mu_1+\mu_2}}
        $%
        }

        \parbox[t]{\dimexpr\linewidth-2em}{%
        $\displaystyle
        q_1^{t+1}(\hat{r}) \gets
        \sum_{z',x'} p^t(\hat{r}|z',x')\,p(z',x')
        $%
        }

        \parbox[t]{\dimexpr\linewidth-2em}{%
        $\displaystyle
        q_2^{t+1}(z|\hat{r},x) \gets
        \frac{p^t(\hat{r}|z,x)\,p(z|x)}
             {\sum_{z'} p^t(\hat{r}|z',x)\,p(z'|x)}
        $%
        }

        \parbox[t]{\dimexpr\linewidth-2em}{%
        $\displaystyle
        q_3^{t+1}(\hat{r}|z) \gets
        \sum_{x'} p^t(\hat{r}|z,x')\,p(x'|z)
        $%
        }

        \parbox[t]{\dimexpr\linewidth-2em}{%
        $\displaystyle
        q_4^{t+1}(r|\hat{r})\! \gets \!
        \frac{\sum_{z',x'} p^t(\hat{r}|z',x')p(z'\!,x'\!,r)}
             {\sum_{z''\!,x''} \!p^t(\hat{r}|z''\!,x'')p(z''\!,x'')}
        $%
        }

      }
    }

    Record converged $\{p_i^T (\hat{r} | z, x)\}$ and its objective value
    $g_i = g(\{p_i^T (\hat{r} | z, x)\})$ \;
}

Output: $\{p^T (\hat{r} | z, x)\}$ corresponding to $\min_i g_i$ \;

\end{algorithm}

In steps 6 and 7 of Algorithm \ref{alg: MI problem}, we use $D(\cdot||\cdot)$ to denote the Kullback-Leibler divergence. More precisely, $D(\{p(r|z,x)\} || \{q_4(r|\hat{r})\}) = \sum_r p(r|z,x) \log \frac{p(r| z, x)}{q_4(r|\hat{r})}$.
The update equations are obtained following steps entirely analogous to those in Sections \ref{subsection: an implicit solution} to \ref{subsection: iterative algorithm}.
To arrive at the exponent in Steps 6 and 7 in the KL-Divergence form, we directly follow steps (26–28) in \cite{tishby2000} after computing the partial derivative of the Lagrangian.
Unlike in the expected distortion problem, this exponent depends on the distribution of $\hat{r}$.
Ultimately, the structure of mutual information introduces an additional independent distribution, $q_4(r \mid \hat{r})$, and consequently, an additional update step.
As the mutual information problem is not convex,
for each $i\in \{1,...., N_j\}$ the solution $p_i^T$ will correspond to a local minimum of $g$. So, we carry out
the BA-style algorithm with multiple random initialisations to increase the likelihood of locating the global minimum.

\section{Numerical Validation of the Blahut-Arimoto-style Procedure} \label{sec: numerical validation of BA}
In this section, we demonstrate the convergence of the Blahut--Arimoto algorithm for the expected distortion problem (Algorithm \ref{alg: BA distortion}), and mutual information problem (Algorithm \ref{alg: MI problem}), and the effectiveness of our procedure in tracing out the $\mathsf{D}(\epsilon, \delta)$ and $\mathsf{I}(\epsilon, \delta)$ curves.
$\mathsf{D}(\epsilon, \delta)$ is the minimum possible expected distortion, subject to privacy constraints parametrised by $\epsilon$ and $\delta$, as in \eqref{eq: optimisation problem new vars}.
$\mathsf{I}(\epsilon, \delta)$ is the maximum possible mutual information subject to the same constraints.\footnote{Strictly speaking, the curve obtained using Algorithm \ref{alg: MI problem} is a lower bound on $\mathsf{I}(\epsilon, \delta)$, as the problem is not convex.}
In our numerical experiments, we take all variables as binary, fix the joint distribution $\{p(r,z,x)\}$ from Table \ref{table: joint dist},
\begin{table}[t] 
\caption{Joint distribution $\{p(z, x, r)\}$ for numerical experiments.}
\centering
\setlength{\tabcolsep}{4pt} % adjust column spacing
\renewcommand{\arraystretch}{1.2} % adjust row spacing
\begin{tabular}{c | cc | cc}
\toprule
\multirow{2}{*}{$z$} & \multicolumn{2}{c|}{$x=0$} & \multicolumn{2}{c}{$x=1$} \\
 & $r=0$ & $r=1$ & $r=0$ & $r=1$ \\
\midrule
0 & $0.024$ & $0.203$ & $0.228$ & $0.013$ \\
1 & $0.063$ & $0.228$ & $0.203$ & $0.038$ \\
\bottomrule
\end{tabular}
\vspace{1mm}
\label{table: joint dist}
\end{table}
and use the Hamming distortion function.

\subsection{Expected Distortion Problem}
\begin{figure}[t]
    \centering  \includegraphics[scale=0.414]{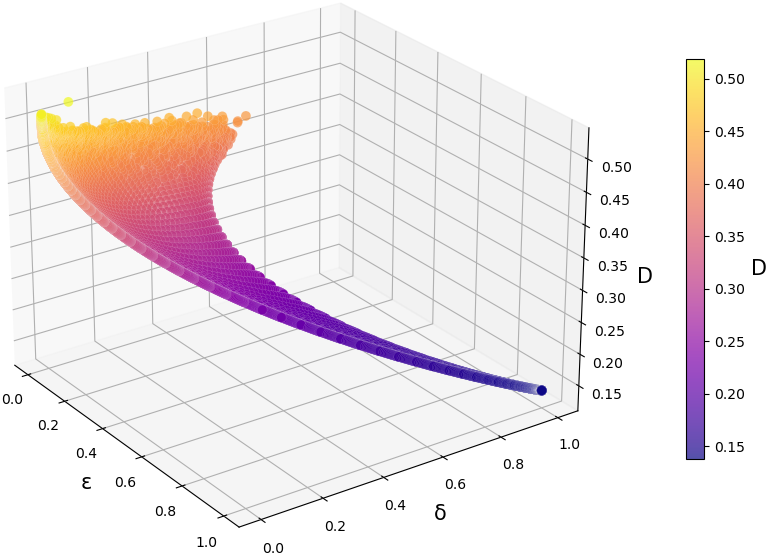}
    \caption{Minimum expected distortion, $\mathrm{D}(\epsilon, \delta)$ against $\epsilon$ and $\delta$.}
    \label{fig: curve}
\end{figure}

Following the curve tracing procedure outlined in Section \ref{section: solution overview}, with the Blahut-Arimoto-style update steps from Algorithm \ref{alg: BA distortion}, we generate the $\mathsf{D}(\epsilon, \delta)$ curve in Figure \ref{fig: curve}.
The resulting surface is convex.
Additionally, fixing either $\epsilon$ or $\delta$ and increasing the other consistently results in lower expected distortion, as one might expect.

\begin{figure}[t]
    \centering  \includegraphics[scale=0.34]{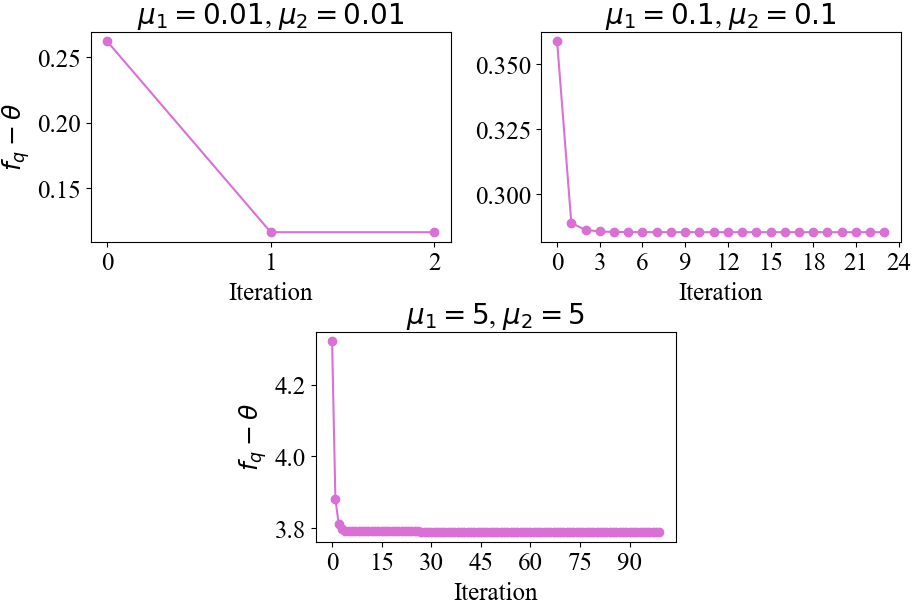}
    \caption{Convergence of Algorithm \ref{alg: BA distortion} for different $\mu_1, \mu_2$ pairs.}
    \label{fig: convergence}
\end{figure}

Figure \ref{fig: curve} indicates that, for each point, the Blahut-Arimoto-style algorithm is converging as expected. 
To verify this further, we examine the convergence of a small set of points: $(\mu_1, \mu_2) = (0.01, 0.01)$, $(\mu_1, \mu_2) = (0.1,0.1)$, and $(\mu_1, \mu_2)= (5,5)$.
The algorithm minimises $f_q$ in a series of update steps. Each set of update steps 5-9 is an iteration.
Recall from \eqref{eq: apply lemma min} and \eqref{eq: apply lemma rearrange} that $\theta$ is constant throughout this process. Thus, we track the progress of $f_q - \theta$ with increasing iterations.
Convergence is illustrated by Figure \ref{fig: convergence}.
When $\mu_1$ and $\mu_2$ are very small, the optimisation boils down to minimising the expected distortion between $\hat{R}$ and $R$.
In this case, the solution converges immediately.
As the values of the Lagrange multipliers increase, the mutual information terms carry greater weight, leading to a slower convergence and an increased number of iterations.
In any case, convergence is consistently fast and the number of required iterations $T$ is reasonably low.

The final solutions, $\{p(\hat{r} \mid  z, x)\}$, in each of the three cases are given in Table \ref{table: cond dists}.
\begin{table}[ht]
\caption{Conditional distributions $\{p(\hat{r} \mid z, x)\}$ for different Lagrange variable pairs.}
    \centering
    \subfloat[\mbox{$(\mu_1,\mu_2) = (0.01,0.01)$}]{%
        \begin{tabular}{cc | cc}
            \toprule
            $z$ & $x$ & $\hat{r}=0$ & $\hat{r}=1$ \\
            \midrule
            0 & 0 & 0 & 1 \\
            0 & 1 & 1 & 0 \\
            1 & 0 & 0 & 1 \\
            1 & 1 & 1 & 0 \\
            \bottomrule
        \end{tabular}
    }
    \hfill
    \subfloat[$(\mu_1,\mu_2) = (0.1, 0.1)$]{%
        \begin{tabular}{cc | cc}
            \toprule
            $z$ & $x$ & $\hat{r}=0$ & $\hat{r}=1$ \\
            \midrule
            0 & 0 & 0.041 & 0.959 \\
            0 & 1 & 0.975 & 0.025 \\
            1 & 0 & 0.143 & 0.857 \\
            1 & 1 & 0.887 & 0.113 \\
            \bottomrule
        \end{tabular}
    }
    \hfill
    \subfloat[$(\mu_1,\mu_2) = (5, 5)$]{%
        \begin{tabular}{cc | cc}
            \toprule
            $z$ & $x$ & $\hat{r}=0$ & $\hat{r}=1$ \\
            \midrule
            0 & 0 & 0.567 & 0.433 \\
            0 & 1 & 0.593 & 0.407 \\
            1 & 0 & 0.367 & 0.630 \\
            1 & 1 & 0.381 & 0.619 \\
            \bottomrule
        \end{tabular}
    }
    \vspace{1mm}
    \label{table: cond dists}
\end{table}
When $\mu_1=\mu_2=0.01$, the focus is entirely on minimising the expected distortion between $\hat{R}$ and $R$.
From Table \ref{table: joint dist}, we know that $R \neq X$ is likely. Therefore, the first solution simply dictates that $\hat{R} \neq X$.
When $\mu_1$ and $\mu_2$ grow to $0.1$, the same logic applies, but noise must be added to satisfy the privacy constraints.
Finally, when $(\mu_1, \mu_1) = (5,5)$, $X$ and $\hat{R}$ are less correlated. Privacy has been prioritised, and a lot of noise is introduced.

\subsection{Mutual Information Problem}
\begin{figure}[t]
    \centering  \includegraphics[scale=0.475]{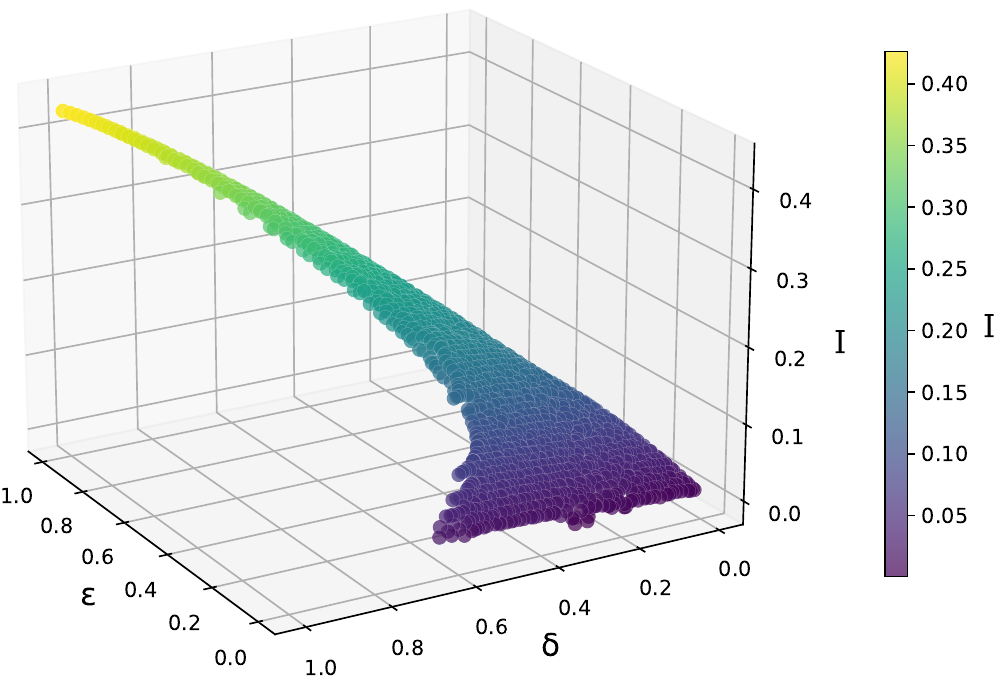}
    \caption{Lower bound on the maximum mutual information $\mathrm{I}(\epsilon, \delta)$ against $\epsilon$ and $\delta$.}
    \label{fig: MI curve} 
\end{figure}

To generate the $\mathsf{I}(\epsilon, \delta)$ curve (or more accurately, its lower bound), we follow the curve tracing procedure from Section \ref{section: procedure}, with the Blahut--Arimoto steps from Algorithm \ref{alg: MI problem}.
The implementation of the procedure may be affected by the non-convex nature of the mutual information problem.
More precisely, the resulting curve may be sparse, with visible gaps in the solution plane.
This is because the dual function is not tight and the inner minimisation in \eqref{eq: inner min} may have many local minima.
Thus, the mapping from dual to primal solutions may not be smooth and continuous.
To generate a continuous surface, a time-sharing scheme can be applied to interpolate between existing points.
A conditional distribution is generated by selecting three points at random and forming a random convex combination of their distributions.\footnote{The new distribution is given by $\{p(\hat{r}| z, x)\} = \lambda_1 \{p_1(\hat{r}| z, x)\} + \lambda_2 \{p_2(\hat{r}| z, x)\} + \lambda_3 \{p_3(\hat{r}| z, x)\}$, where $\lambda$ coefficients are positive and sum to $1$.
The appearance of the surface is improved by biasing time-sharing samples $p_1, p_2, p_3$ away from densely populated regions of the curve.}
This distribution defines a feasible point, which is retained if it lies on the upper envelope of the 
surface constructed from the existing points (i.e., the boundary of our current $\mathsf{I}(\epsilon, \delta)$ approximation, not necessarily the optimal surface).
The result of this process is displayed in Figure \ref{fig: MI curve}.

\section{Adaptive Privacy Scheme}
\label{section: sequential solution procedure}
Our procedure to solve the privacy problem given in Section \ref{section: problem setup} is as follows.
A data handler receiving $m$ sequential requests for information should respond by applying Algorithm~\ref{alg: sequential release}.
\begin{algorithm}[t]
\caption{Sequential Data Release Strategy via Blahut--Arimoto}
\label{alg: sequential release}
\SetAlgoLined
\DontPrintSemicolon

\For{each $k=1,\dots,m$}{
  Input: Request for $R_k$

  Assign $\hat{R} \leftarrow \hat{R}_k$, $Z \leftarrow (\hat{R}_1,\dots,\hat{R}_{k-1})$, $R \leftarrow R_k$, $\delta  \leftarrow \delta_k$, $\epsilon \leftarrow \epsilon_k$ \;

  Define a grid $\{(\mu_{1,j},\mu_{2,j})\}_{j\in J}$ of Lagrange multipliers \;

  \For{each $j\in J$}{
    Run the Blahut--Arimoto algorithm at $(\mu_{1,j},\mu_{2,j})$ to obtain $\{p_j^T(\hat{r}\mid z,x)\}$ \;
    Compute the corresponding triple
    $\mathsf{U}_j \leftarrow - U(\hat{R}, R)$, $\epsilon_j \leftarrow I(\hat{R};X)$, $\delta_j \leftarrow I(\hat{R}, Z;X)$
  }
  Collect all points $(\mathsf{U}_j, \epsilon_j, \delta_j)$ as a coarse approximation to the entire curve $\mathsf{D}(\epsilon,\delta)$ or $\mathsf{I}(\epsilon,\delta)$\; 

Use bisection search on $(\mu_1, \mu_2)$ within the grid to find the pair and corresponding $\{p^T(\hat{r}\mid z,x)\}$ giving the target $(\epsilon, \delta)$; if no such pair exists, locate it with time-sharing by iteratively adjusting convex combinations of nearby conditional distributions until convergence to $(\epsilon,\delta)$\;

  Compute $\hat{R}$ according to $\{p^T(\hat{r}\mid z,x)\}$ and set $\hat{R}_k \leftarrow \hat{R}$ \;

  Output: Release the value of $\hat{R}_k$
}
\end{algorithm}
The Blahut--Arimoto algorithm called in steps 6 and 10 is Algorithm \ref{alg: BA distortion} for the expected distortion problem, or Algorithm \ref{alg: MI problem} for the mutual information problem.

The resulting data release is optimal with respect to the expected distortion problem. In contrast, no such guarantee holds for the mutual information problem, which is non-convex. Consequently, the Blahut--Arimoto algorithm is only guaranteed to reach a local minimum, possibly yielding a suboptimal solution. To improve the likelihood of identifying the optimal solution, we employ multiple initializations.
The Blahut--Arimoto step maintains an efficient run time.
In fact, we can characterise its complexity as $O( T  \,| \mathcal{\hat{R}}| |\mathcal{Z}| | \mathcal{X}|\, N_j)$, where $N_j=1$ for the expected distortion problem.
Recalling that $\mathcal{Z}$ represents $\hat{\mathcal{R}}^{k-1}$, we remark that the complexity increases with increasing $k$.
As the total information leakage of the adaptive privacy scheme should be limited, we expect $m$ to remain within a manageable range.
As a result, the total complexity will not increase without bound.

\section{Empirical Validation on Real Data}
Section \ref{sec: numerical validation of BA} demonstrates feasibility and convergence of the BA-style algorithm on a synthetic binary example.
In this section, we evaluate the adaptive privacy scheme of Section \ref{section: sequential solution procedure} on a real dataset, and compare it against suitable baselines.
Since no existing privacy mechanism directly addresses our setting,
there is no direct state-of-the-art baseline for our problem. 
We therefore compare against two natural benchmarks.
First,
a non-adaptive scheme, which isolates the value of conditioning on previous releases.
Second, a symmetric-channel mechanism, which isolates the performance of our optimisation scheme.

\subsection{Dataset and Experimental Setup}
We carry out experiments on the UCI Adult dataset, also known as the Census Income dataset \cite{uci_adult}, 
from which we construct an empirical distribution over discretised records. 
Specifically, we extract three attributes: education, income, and age, which we discretise into 4, 2, and 4 bins respectively. 
We define the private variable as $X=(E,J,A)$,
where $E$, $J$, and $A$ represent the discretised education, income, and age variables.
Hence, $|\mathcal X|=32$. 
Throughout, we focus on the expected distortion problem, taking the cost function to be Hamming distortion.

\subsection{Sequential Data Release Experiment} \label{sec: experiment 1}
First, we consider a set of four sequential requests, where $(R_1, R_2, R_3, R_4)= (E, J, A, E)$.
In other words, the data handler receives requests for education, income, and age attributes in turn, followed by a repeated request for the education attribute.

For each request, we apply Algorithm \ref{alg: sequential release} to find the optimal release distribution under the expected distortion formulation, i.e., the solution to \eqref{eq: optimisation problem}.
We call this the \emph{adaptive mechanism}.
To assess the effect of the collusion constraint, we compare with the solution to the \emph{non adaptive problem}, by which we mean
\begin{align*}
\min_{\{p(\hat{r}_k \mid x)\}} \quad & - U(\hat{R}_k, R_k)  \\
\text{s.t.} \quad & I(\hat{R}_k; X) \leq \epsilon_k,
\end{align*}
and whose output we call the \emph{non-adaptive mechanism}.
The non-adaptive problem is solved separately for each release and does not depend on previous outputs.\footnote{To solve the non-adaptive problem, we use Algorithm~\ref{alg: sequential release}, setting $Z$ as a dummy variable rather than $\hat{R}^{k-1}$.}
We nevertheless evaluate its cumulative leakage as $I(\hat{R}^k; X)$, allowing direct comparison with the adaptive mechanism.

\begin{figure*}[t]
    \centering  
    \includegraphics[scale=0.44]{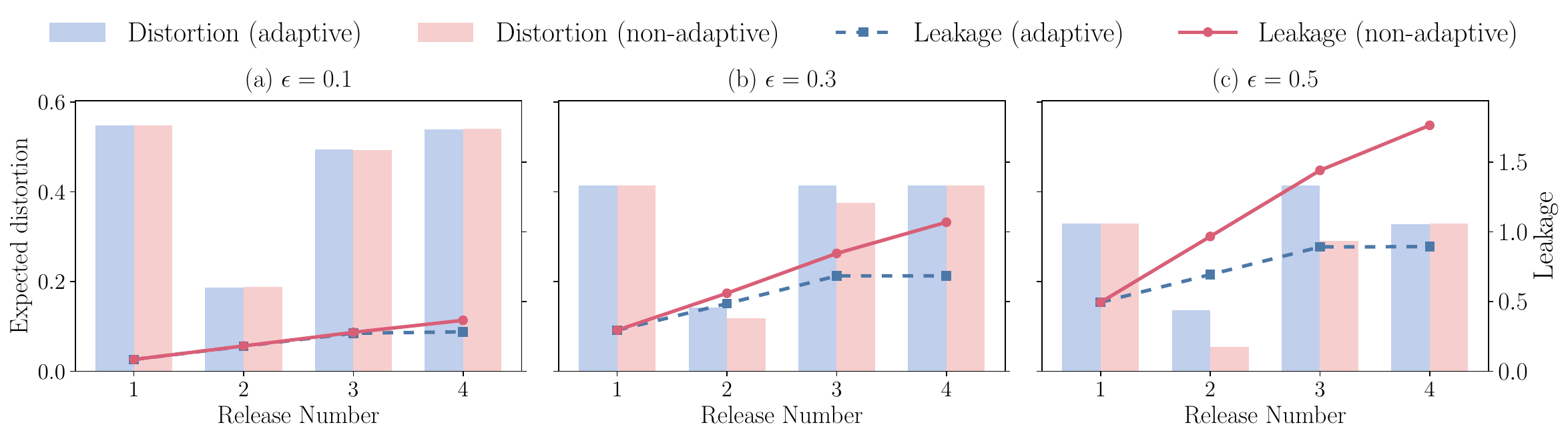}
    \caption{Expected Hamming distortion (bars) and cumulative leakage (lines) across four sequential requests for education, income, age, and education. Results are shown for uniform $\epsilon_k = \epsilon \in \{0.1, 0.3, 0.5\}$, $\delta_k = 0.2(k-1)+\epsilon$.}
    \label{fig: 4 release} 
\end{figure*}

We use a uniform privacy budget across releases, i.e., $\epsilon_1 = \epsilon_2 = \epsilon_3 = \epsilon_4 = \epsilon$, and test $\epsilon \in \{0.1, 0.3, 0.5\}$.
In each case, $\delta_k$ is chosen to increase linearly with $k$ according to $\delta_k = 0.2 (k-1) + \epsilon$.
Figure~\ref{fig: 4 release} shows the resulting cumulative leakage and distortion across releases.

Since the non-adaptive problem is equivalent to \eqref{eq: optimisation problem} for the first release, distortion and cumulative leakage are identical when $k=1$.
For subsequent releases, the cumulative leakage grows more quickly in the non-adaptive case.
The effect is more pronounced as $\epsilon$ grows, since larger $\epsilon$ means $X$ shares more information with previous releases.
There are two reasons for the increased growth in cumulative leakage.
First, since the second constraint has been dropped, more information may be shared in total.
This benefits expected distortion, which is visibly smaller for the non-adaptive mechanisms on the second and third releases when $\epsilon \in \{0.3, 0.5\}$.
Second, the non-adaptive mechanism cannot reuse information from previous releases.
This means that any information shared to reduce distortion must be independently generated across releases, and thus may be combined by an adversary to reveal more information than intended.
This effect is prominent in the fourth release.
Since $R_4 = R_1$, the adaptive mechanism can reuse $\hat{R}_1$ in full, matching the expected distortion achievable by a non-adaptive mechanism, without introducing any extra information.
Hence, across $\epsilon$ values, we observe that when $k=4$, the non-adaptive mechanism suffers significant additional cumulative leakage without any utility advantage.

\subsection{Repeated Request Trade-off Curves} \label{sec: experiment 2}
We have found that adaptivity is particularly valuable when a request contains information that has already been released. 
To investigate this effect more directly, we next consider a setting in which the same attribute is requested twice: $(R_1, R_2) = (E, E)$.
In the experiment, the first release has already been obtained using Algorithm~\ref{alg: sequential release} with $\epsilon_1=\delta_1=0.3$.
The focus is the privacy-utility trade-off of the second release.
For this release, we vary the individual privacy budget $\epsilon_2 \in [0.3, H(E)]$ and compare the resulting trade-off curves under four mechanisms: the adaptive mechanism with $\delta = \epsilon$ and with $\delta = \infty$, the non-adaptive mechanism, and a symmetric-channel baseline.  
We choose a symmetric channel because it is simple and interpretable, and  serves as a generic benchmark that does not explicitly optimise the privacy–utility-collusion trade-off.
Since Algorithm \ref{alg: sequential release} is designed to solve the optimisation problem in \eqref{eq: optimisation problem}, it should perform at least as well as any agnostic mechanism. 
The comparison works to empirically validate our optimisation procedure.

For the adaptive mechanisms, we obtain the second release distribution 
from Algorithm \ref{alg: sequential release}, which solves \eqref{eq: optimisation problem}.
For the non-adaptive mechanism, the second release is optimised without conditioning on $\hat{R}_1$, as in the previous subsection. 
We remark that the objective function and first constraint do not depend on $\hat{R}_1$.
Therefore, setting $\delta=\infty$ which effectively removes the second constraint, in theory, reduces the optimisation problem to the non-adaptive problem. 
However, we will observe that the numerical solutions can still differ because the Blahut-Arimoto-style updates may guide the optimisation toward different minima.
Finally, the symmetric-channel baseline releases the correct education value with probability $p$, and each incorrect value with equal probability.

\begin{figure}[ht]
    \centering  \includegraphics[scale=0.44]{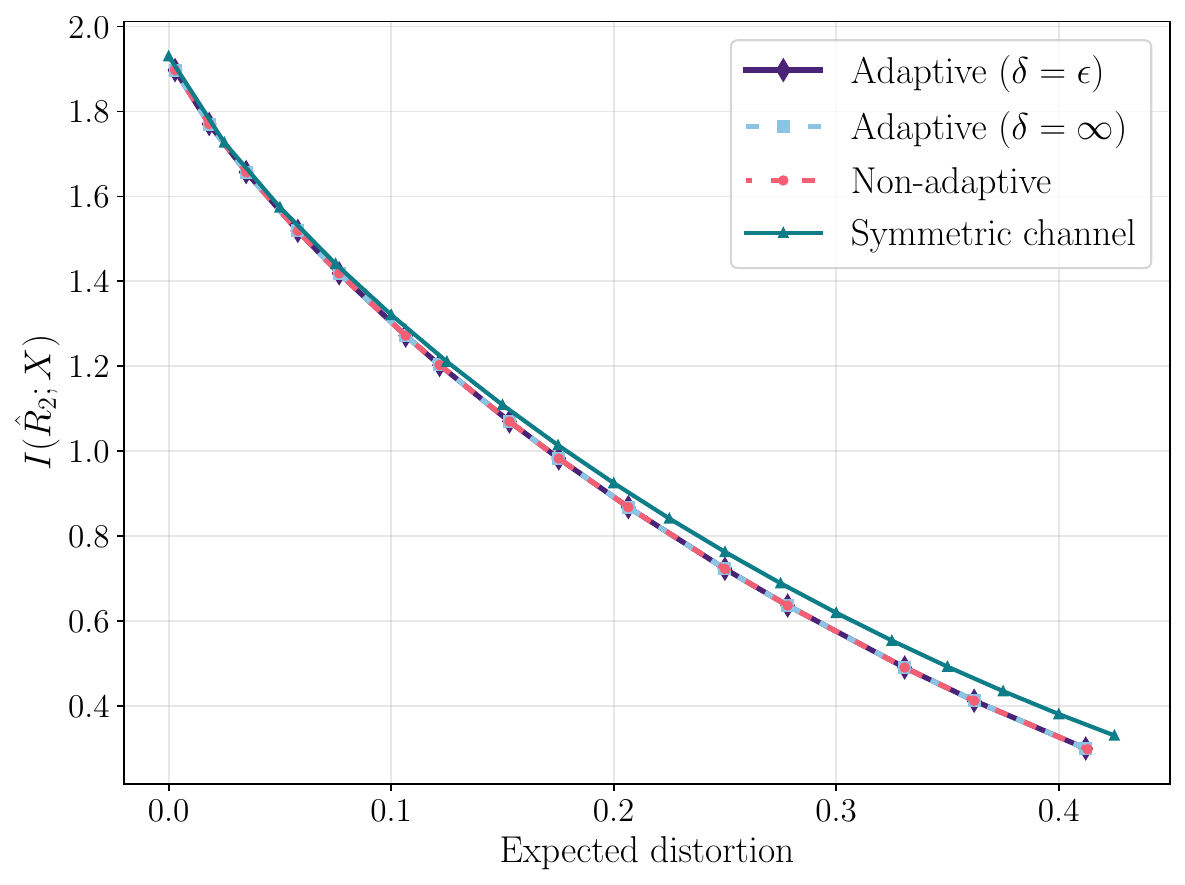}
    \caption{Individual leakage $I(\hat{R}_2;X)$ against expected Hamming distortion for a repeated education request, with $\epsilon_1=\delta_1=0.3$ and $\epsilon_2$ varied.}
    \label{fig: tradeoff 1} 
\end{figure}

\begin{figure}[ht]
    \centering  \includegraphics[scale=0.44]{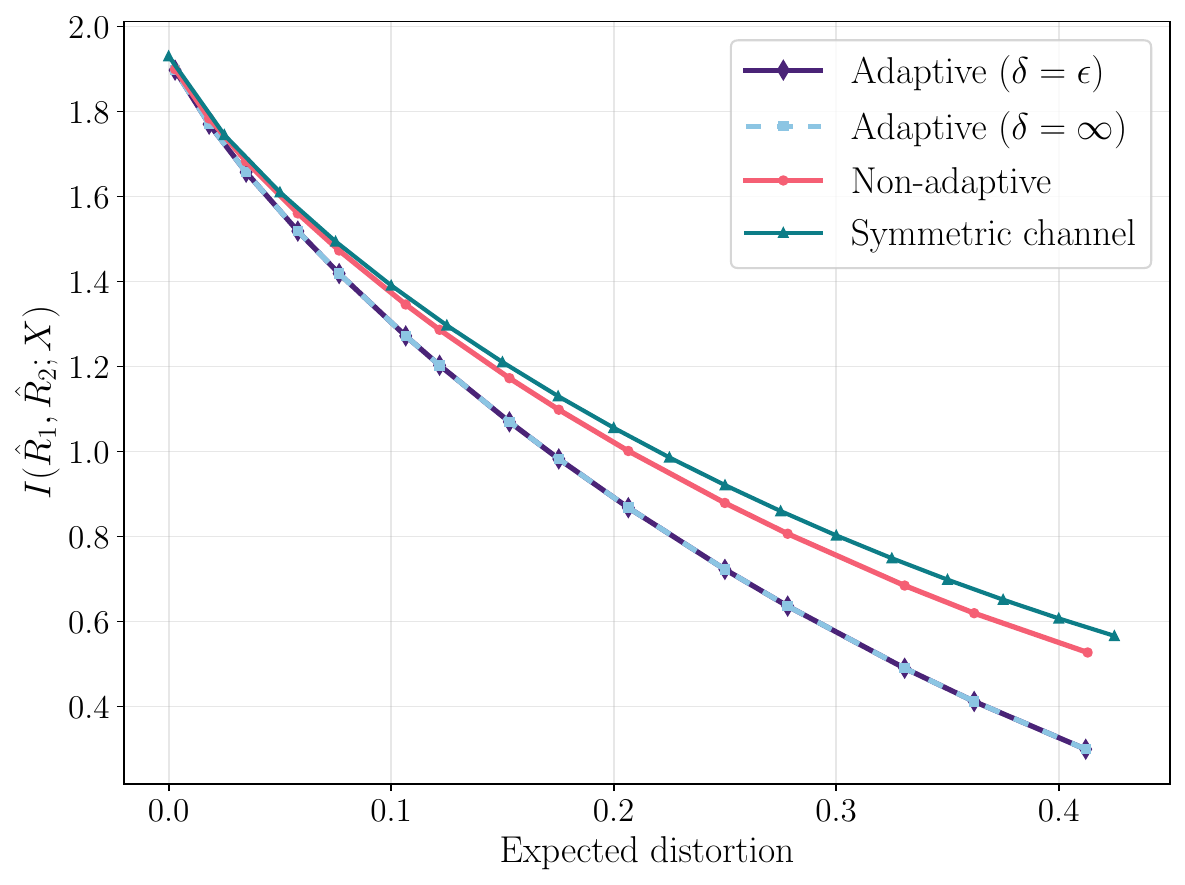}
    \caption{Cumulative leakage $I(\hat{R}_1, \hat{R}_2;X)$ against expected Hamming distortion for a repeated education request, with $\epsilon_1=\delta_1=0.3$ and $\epsilon_2$ varied.}
    \label{fig: tradeoff 2} 
\end{figure}

We plot the expected distortion against both the individual leakage $I(\hat{R}_2;X)$ and the cumulative leakage $I(\hat{R}_1,\hat{R}_2;X)$. 
In doing so, we separately compare the individual privacy-utility trade-off of the second release, and the equivalent trade-off under full collusion.
Figure~\ref{fig: tradeoff 1} shows the former, and Figure~\ref{fig: tradeoff 2} the latter.

The first observation is that in Figure~\ref{fig: tradeoff 1}, the trade-off curves coincide for the adaptive and non-adaptive mechanisms, meaning all three procedures achieve the same utility for each individual leakage value. 
Since the second request repeats the first, the adaptive mechanism can use $\hat{R}_1$ for free as far as the second constraint is concerned. 
In particular, if $\hat{R}_2$ includes $\hat{R}_1$, $I(\hat{R}_1 \hat{R}_2; X) = I(\hat{R}_2; X)$, so it is natural that the adaptive mechanism may bypass the restrictions of the second constraint, even when $\delta_2=\epsilon_2$.
Thus, in this case, using the constrained adaptive mechanism does not incur any utility loss.
The second observation is that the curve traced by the symmetric channel is dominated by those obtained using Algorithm~\ref{alg: sequential release},
demonstrating that our proposed procedure provides a systematic improvement over this baseline.

Things change slightly when cumulative leakage is plotted against expected distortion.
In Figure~\ref{fig: tradeoff 2}, the non-adaptive mechanism can no longer match the performance of the adaptive mechanisms.
This behaviour is expected because a non-adaptive mechanism must generate $\hat{R}_2$ without access to $\hat{R}_1$.
Consequently, for $\hat{R}_2$ to be useful, it must be that $I(\hat{R}_1 \hat{R}_2; X) > I(\hat{R}_2; X)$.
More interestingly, the performance of the adaptive mechanism appears to be invariant to the choice of $\delta_2$.
At first glance, this is somewhat surprising; if $\delta_2 = \infty$, the second constraint is inactive and $\hat{R}_2$ need not make use of $\hat{R}_1$.
Nevertheless, the cumulative leakage matches that of when both constraints are active.
The result points to an interesting feature of the Blahut-Arimoto-style procedure, suggesting that it naturally favours solutions that reuse information.
In other words, the history-aware update equations seem to push solutions towards minima that exploit prior releases, even when such reuse is not enforced by the constraints.

\subsection{Discussion of Findings}
The experiments from Sections \ref{sec: experiment 1} and \ref{sec: experiment 2}
each demonstrate that the proposed procedure controls cumulative leakage in settings where independently designed mechanisms can exceed the intended leakage under collusion. 
This comparison is particularly relevant because most privacy-utility frameworks are formulated as single-release optimisation problems \cite{Dwork2010,Riboni2012,Erdogdu2015,Hasan2017,Dong2024,SHMUELI2015, LocationTrace}, or as sequential schemes in which mechanisms are designed without conditioning on previous outputs \cite{Zamani2021}. 
Applying non-adaptive mechanisms at each request yields the non-adaptive baseline considered in this section.

In particular, the repeated-request experiments in both sections demonstrate that complete adaptivity can improve privacy without sacrificing utility when requested information overlaps with previous releases. 
In these cases, the mechanism can reuse information that has already been revealed, reducing cumulative leakage while maintaining the same expected distortion. 
These observations suggest that conditioning on disclosure history is particularly valuable when privacy guarantees must remain valid under collusion by multiple parties.

The trade-off curves additionally show that the solutions produced by Algorithm~\ref{alg: sequential release} dominate the symmetric-channel baseline. 
This empirical result is consistent with the optimality result established in Section \ref{subsection: iterative algorithm}, and demonstrates the advantage of the Blahut-Arimoto-style procedure over more simply designed mechanisms.

\section{Connections to Machine Learning}

As discussed in Section \ref{section: solution overview}, the optimisation problem in our adaptive privacy framework can be seen as a generalisation of the information bottleneck problem. As a result, it also finds applications in the context of machine learning. The information bottleneck has proven influential in deep learning literature, inspiring a great deal of neural network algorithms since its conception in \cite{tishby2000}.
The IB method balances compression and relevance, and is a way of limiting generalisation errors in deep learning \cite{IB2023}.
Consider a set of training data $X$ and a task $Y$. The IB method generates an output $T$ that minimises
\begin{align*}
    I(T; X) - \beta I(T; Y).
\end{align*}
The second term promotes relevancy or utility, whilst the first promotes compression, ensuring that superfluous information is discarded.
It is worth noting that in deep learning, conditional distributions are not generally optimised over directly, but are parametrised and trained via gradient descent.
Also, mutual information terms tend to be approximated by more tractable functions.
Despite this, the exact solution to the IB problem is useful as a benchmark, and as a tool to gain insight into the fundamental trade-offs between compression and relevance.
In this section, we propose an extension to this idea, where the IB method is adapted to accommodate sequential tasks.
In particular, we will consider the application of progressive neural networks.

Progressive neural networks (PNNs) \cite{PNNs2022} provide responses to sequential tasks by creating a new neural network (or column) for each.
Each column depends on previous columns via lateral connections.
Specifically, each hidden layer receives inputs from the previous layer of its own column as well as from the corresponding hidden layers of all previously trained columns.
As past responses to tasks are stored rather than overwritten, PNNs avoid the problem of catastrophic forgetting.
This comes at the cost of increased memory usage, making the scalability of PNNs a significant research challenge.
We address this challenge in the following framework.

Let $Y_1, \dots, Y_m$ be a set of sequentially received tasks, with $T_1, \dots, T_m$ representing the corresponding outputs in a PNN.
The training data is denoted by $X$.
We let $Y$ and $T$ denote the current task and output respectively, whilst $Z$ represents all previous outputs.
Consider the following choice for $T$:
\begin{align} \label{eq: PNN IB}
    \min_{\{p(t  \mid  z, x ) \} } - I(T; Y) + \mu_1 I(T; X) + \mu_2 I(T, Z; X).
\end{align}
Like the IB method, relevancy is promoted by the $I(T; Y)$ term. The second term, $\mu_1 I(T; X)$, encourages compression of 
$X$, ensuring that irrelevant information is excluded from the output.
The goal of the final term, $ \mu_2 I(T, Z; X)$ is to address the memory issues of PNNs. 
The quantity $(T, Z)$ represents the total information that must be stored after $T$ is generated.
Thus, the last term ensures that extra data only occupies memory if it is beneficial.
The effect of the term is to promote shared information between $Z$ and $T$ where possible.
We assume that redundancies in the outputs can be leveraged for more efficient storage, lowering memory requirements.
Note that if $\mu_1 \geq 1$, the optimal solution will always be $T = \emptyset$.
Therefore, a suitable range for $\mu_1$ is $0 \leq \mu_1 \leq 1$, where $\mu_1 = 0$ corresponds to fully prioritising relevancy, and $\mu_1 = 1$ corresponds to fully prioritising compression.

Since the optimisation problem \eqref{eq: PNN IB} matches the inner minimisation of the machine learning problem dual \eqref{eq: opt prob} without constant terms, we can simply make use of Algorithm \ref{alg: MI problem}.
As weightings $\mu_1$ and $\mu_2$ are the parameters of interest, there is no need for a bisection search to find the particular solution.
Applied sequentially, this procedure can provide insights into the underlying trade-offs between compression, memory, and relevance in progressive neural networks.

\section{Conclusion} \label{section: collusion}

In this work, we proposed a problem formulation for handling multiple sequential data requests while balancing privacy and utility, accounting for the possibility of collusion.
We allow utility to be defined according to expected distortion or mutual information.
We outlined a procedure to adaptively release data and trace the three-dimensional distortion-privacy-collusion curve $\mathsf{D}_k(\epsilon_k, \delta_k)$, or information-privacy-collusion curve $\mathsf{I}_k(\epsilon, \delta)$, for each party $k$, analogous to 
the existing approach for the rate-distortion problem.
This involves converting the problem to its dual, fixing a grid of Lagrange multipliers, and solving the inner minimisation with the Blahut-Arimoto-style algorithm.
We provided the steps for such an algorithm and showed that it ensures convergence to the appropriate value.
As in the rate-distortion problem, the solution to a specific primal problem, and thus the distribution for data release, can be found through a bisection search over the Lagrange multipliers.
From this, we proposed a sequential data release algorithm to systematically respond to requests, maximising utility whilst adhering to individual and collusion-based privacy constraints.

By means of a numerical example, we demonstrated the feasibility of our procedure and confirmed the convergence of the Blahut--Arimoto algorithm for the expected distortion problem.
We further validated the framework on the UCI Adult dataset, comparing the proposed sequential algorithm with non-adaptive and symmetric-channel baselines. 
The results highlighted that in some cases, conditioning on previous releases substantially reduces cumulative leakage under collusion, without sacrificing utility.

Finally, we discussed a possible extension to the application of progressive neural networks.
We outlined a relevant optimisation problem and
remarked that its solution can be found using the Blahut--Arimoto algorithm designed for our mutual information adaptive privacy problem.

Further work could extend our framework to enable more general guarantees against collusion.
For example, constraints could be chosen to limit the data leakage should $n$ out of $m$ parties collude.
Even more generally, one could limit leakage should certain subsets of the parties collude. 
The subsets can be chosen based on knowledge of the parties and their relationships.
The $k$th optimisation problem would be
\begin{align*}
\min_{\{p(\hat{r}_k \mid \hat{r}^{k-1}, x)\}} \quad & - U(\hat{R}_k, R_k)  \\
\text{s.t.} \quad 
& I(\hat{R}_k; X) \leq \epsilon_k, \\
& I\big(\hat{R}_k, \hat{R}_{S_k^i}; X\big) \leq \delta_k^i,
  \\& S_k^i \subseteq \{1,\dots,k-1\},\; i=1,\dots,\sigma .
\end{align*}
Here, $\sigma$ denotes the number of subsets of concern that are formed from the first $k$ parties and contain party $k$.  
Each $S_k^i$ is one such subset, consisting of indices in $\{1,\dots,k-1\}$ (i.e., excluding $k$).  
We write $\hat{R}_{S_k^i} := \{\hat{R}_j : j \in S_k^i\}$ for the set of data releases associated with $S_k^i$.
Our adaptive privacy scheme, including the Blahut--Arimoto algorithm, can be extended to this case.

\bibliographystyle{IEEEtran}
\bibliography{references}  

\end{document}